\documentclass[preprint,sort&compress,review,11pt]{elsarticle}
 \usepackage[top=1in, bottom=1in, left=1in, right=1in]{geometry}

\usepackage{graphicx}
\usepackage{subfig}
\usepackage{graphicx}

\usepackage{xcolor}
\usepackage{color,soul}

\usepackage{amsmath}
\usepackage{amssymb}
\usepackage{amsthm}
\usepackage{stmaryrd}
\usepackage{algorithm}
\usepackage{algpseudocode}
\usepackage{bbm}
\usepackage{bm}
\usepackage{mathtools}
\usepackage{tikz}

\usepackage{comment}
\usepackage{url}
\usepackage{listings}
\usepackage[colorlinks=true]{hyperref}
\usepackage{array}
\usepackage{multirow}
\usepackage{enumitem}
\newcolumntype{P}[1]{>{\centering\arraybackslash}m{#1}}
\newcolumntype{L}[1]{>{\raggedright\let\newline\\\arraybackslash\hspace{0pt}}m{#1}}
\newcolumntype{C}[1]{>{\centering\let\newline\\\arraybackslash\hspace{0pt}}m{#1}}
\newcolumntype{R}[1]{>{\raggedleft\let\newline\\\arraybackslash\hspace{0pt}}m{#1}}

\usepackage{lineno}
\newcommand*\patchAmsMathEnvironmentForLineno[1]{%
  \expandafter\let\csname old#1\expandafter\endcsname\csname #1\endcsname
  \expandafter\let\csname oldend#1\expandafter\endcsname\csname end#1\endcsname
  \renewenvironment{#1}%
     {\linenomath\csname old#1\endcsname}%
     {\csname oldend#1\endcsname\endlinenomath}}%
\newcommand*\patchBothAmsMathEnvironmentsForLineno[1]{%
  \patchAmsMathEnvironmentForLineno{#1}%
  \patchAmsMathEnvironmentForLineno{#1*}}%
\AtBeginDocument{%
\patchBothAmsMathEnvironmentsForLineno{equation}%
\patchBothAmsMathEnvironmentsForLineno{align}%
\patchBothAmsMathEnvironmentsForLineno{flalign}%
\patchBothAmsMathEnvironmentsForLineno{alignat}%
\patchBothAmsMathEnvironmentsForLineno{gather}%
\patchBothAmsMathEnvironmentsForLineno{multline}%
}

\usepackage[colorinlistoftodos]{todonotes}
\definecolor{lightblue}{rgb}{.90,.95,1}
\definecolor{darkgreen}{rgb}{0,.5,0.5}

\definecolor{lightgreen}{rgb}{.90,1,0.90}

\usepackage{changes}

\definechangesauthor[color=orange]{hx}
\definechangesauthor[color=blue]{cm}
\makeatletter
\setremarkmarkup{\todo[color=Changes@Color#1!20,size=\scriptsize]{#1: #2}}
\makeatother

\providecommand\bnabla{\boldsymbol{\nabla}}

\journal{Computers and Fluids}

\begin{document}

\begin{frontmatter}

\title{Flows Over Periodic Hills of Parameterized Geometries:\\  A Dataset for Data-Driven Turbulence Modeling From Direct Simulations}

\author{Heng Xiao\corref{mycor}}%
\ead{hengxiao@vt.edu}
\cortext[mycor]{Corresponding author}

\author{Jin-Long Wu\fnref{fnjl}}
\ead{jinlong@vt.edu}
\address{Kevin T. Crofton Department of Aerospace and Ocean Engineering, Virginia Tech, Blacksburg, Virginia, USA}

\author{Sylvain Laizet\corref{slcor}}
\address{Department of Aeronautics, Imperial College London, UK}

\fntext[fnjl]{Current affiliation: California Institute of Technology, Pasadena, California, USA.}
\author{Lian Duan\corref{ldcor}}

\address{Department of Mechanical and Aerospace Engineering, The Ohio State University, Columbus, Ohio, USA}

\begin{abstract}
Computational fluid dynamics models based on Reynolds-averaged Navier--Stokes equations with turbulence closures still play important roles in engineering design and analysis. However, the development of turbulence models has been stagnant for decades. 
With recent advances in machine learning, data-driven turbulence models have become attractive alternatives worth further explorations. However, a major obstacle in the development of data-driven turbulence models is the lack of training data. In this work, we survey currently available public turbulent flow databases and conclude that they are inadequate for developing and validating data-driven models. Rather, we need more benchmark data from systematically and continuously varied flow conditions (e.g., Reynolds number and geometry) with maximum coverage in the parameter space for this purpose. To this end, we perform direct numerical simulations of flows over periodic hills with varying slopes, resulting in a family of flows over periodic hills which ranges from incipient to mild and massive separations. We further demonstrate the use of such a dataset by training a  machine learning model that predicts Reynolds stress anisotropy based on a set of mean flow features. We expect the generated dataset, along with its design methodology and the example application presented herein, will facilitate development and comparison of future data-driven turbulence models.
\end{abstract}


\end{frontmatter}

\section{Introduction}
\label{sec:intro}

Although turbulence affects natural and engineered systems from sub-meter to planetary scales, fundamental understanding and predictive modeling of turbulence continue to defy and bedevil scientists and engineers. Turbulent flows are typical multi-scale physical systems that are characterized by a wide range of spatial and temporal scales. When predicting such systems, first-principle-based simulations are prohibitively expensive, and the small-scale processes must be modeled.  For simulating turbulent flows, this is done by solving the Reynolds-Averaged Navier--Stokes (RANS) equations with the unresolved processes represented by turbulence model closures.

While the past two decades have witnessed a rapid development of high-fidelity turbulence simulation methods such as
large eddy simulations (LES), they are still too expensive for practical systems such as the flow around an commercial airplane~\cite{moin97tackling}. It is expect that using LES for engineering design will remain infeasible for decades to come. On the other hand, attempts in combining models of different fidelity levels (e.g., hybrid LES/RANS models) have shown promises, but how to achieve consistencies in the hierarchical coupling of models is still a challenge and a topic of ongoing research.
Consequently, Reynolds-Averaged Navier--Stokes (RANS) equations are still the workhorse tool in engineering computational fluid dynamics for simulating turbulent flows.

It is well known that RANS turbulence models have large model-form uncertainties for a wide range of flows~\cite{xiao2019quantification}, which diminish the predictive capabilities of the RANS-based CFD models. 
Development of turbulence models has been stagnant for decades, which is evident from the fact that currently used turbulence models (e.g., $k$--$\varepsilon$, $k$--$\omega$, and Spalart--Allmaras models~\cite{launder74application,menter94two-equation,spalart92one}) were all developed decades ago despite unsatisfactory performance for many flows. In view of the importance and stagnation of turbulence modeling, NASA Technology Roadmap~\cite{adler2012entry} called for sustained investment in improvements in the modeling of turbulent flows~\cite{national2012nasa}. In particular, NASA CFD 2030 vision~\cite{slotnick2014cfd} identified the development of advanced turbulence model as a key priority.

Recent advances in machine learning techniques enabled researchers to explore data-driven turbulence models as attractive alternatives to traditional algebraic models and Reynolds stress transport models.  Duraisamy et al.~\cite{singh16using,singh16machine} introduced a multiplicative discrepancy field to the source term of the turbulence transport equations.  Xiao and co-workers~\cite{wang2017physics-informed,wu2018physics-informed} built a regression function for the RANS-modeled Reynolds stress discrepancies in terms of mean flow quantities. Ling et al.~\cite{ling2016reynolds} directly built regression functions for the Reynolds stresses with data from DNS databases. Weatheritt and Sandsberg~\cite{weatheritt2016novel,weatheritt2017development} used symbolic regression and gene expression programming for learning the coefficients in algebraic turbulence models.
Schmelzer et al.~\cite{schmelzer2019machine} used deterministic symbolic  sparse regression to infer algebraic stress models and perform a systematic assessment on separated flows. 

While machine learning has repeatedly exceeded expert expectations in business applications from IBM's Jeopardy-winning Watson to Google's World-Champion-beating AlphaGo, their applications in physical modeling, and particularly turbulence modeling, have been hindered by two major obstacles: (1) the difficulty in incorporating domain knowledge of the physical systems (e.g., conservation laws, realizability, energy spectrum), and (2) the lack of publicly available, high-quality training flow databases. In the fields where machine learning has achieved tremendous successes (e.g., computer vision as detailed below), neither of these difficulties exist. This work will primarily discusses the bottleneck of training data for data-driven turbulence modeling.

\subsection{The need for parameterized datasets for data-driven turbulence modeling}

Training datasets play pivoting roles in developing algorithms in machine learning and data-driven modeling. For example, in the field of image classification and computer vision, the dataset ImageNet has greatly facilitated the advancement of machine learning algorithms in the past decades~\cite{krizhevsky12imagenet}. As of now, the ImageNet dataset contains 14 million images hand-annotated with one of the 20,000 labels, which indicates what object is pictured~\cite{deng09imagenet}.

Currently available public datasets are inadequate for training and testing data-driven turbulence models for RANS simulations. For example, the JHU database consists of DNS data of simple flows such as homogeneous isotropic turbulence and plane channels flows of friction Reynolds numbers ranging from $Re_\tau = 180 $ to $Re_\tau = 5200$~\cite{kim1987turbulence,lee15direct}. Despite the high quality and large amounts of the data, such flows are not the focus of RANS modeling. The AGARD dataset~\cite{agard} consists of a number of flows of engineering interests, but the data are meant for validating LES. The NASA Langley turbulence modeling portal~\cite{nasa-database} contains a number of complex flows of interest to the RANS modeling community. This database is unique in that it is intended to serves as \emph{verification} of user-implemented RANS turbulence models rather than for validation efforts. That is, they provided the solution a particular RANS turbulence model should produce rather than the true physical solution, although the data for the latter are often available as well. All the above-mentioned datasets consist of representative flows that are clearly distinct from one another. For example, typical flows include attached boundary layers, separated flows, bluff body wakes, and jets. While such dataset are suitable for validating the implementation of traditional models, they are not suitable for developing data-driven models. In summary, existing high-fidelity simulations (DNS and LES) are performed to probe turbulent flow physics, and the simulations are often performed on representative but sufficiently different configurations. As a result, existing benchmark databases~\cite{nasa-database, ercoftac,agard,jhu-database,li08public} are sparsely scattered in the parameter space, as they were generated for understanding flow physics or validating CFD solvers.

There are two noteworthy exceptions. First, Pinelli et al.~\cite{pinelli10reynolds} performed DNS of fully developed turbulent flows in a square duct at Reynolds numbers ranging from 1400 to 2500. The data are well curated and publicly available at their institutional website~\cite{kti-database}. This flow is of great interest to the turbulence modeling community, because the in-plane secondary flows (recirculation) induced by the imbalance of Reynolds normal stresses cannot be captured by linear eddy viscosity models (the most commonly used models at present). This feature is challenging to capture even with advanced models such as nonlinear eddy viscosity models and Reynolds stress transport models. A shortcoming of this dataset is that the series of flows differ only in Reynolds numbers and not in geometry. It would be a more challenging test if a model trained on flows in square ducts are used to predict flows in rectangular ducts. The same group has performed DNS of flows in rectangular ducts, but the data are not yet publicly available.
Another set of flows that can potentially be used for developing data-driven models is the flows separated from smoothly contoured surfaces. With benchmark data scatted in the NASA database~\cite{nasa-database} and the ERCOFTAC database~\cite{ercoftac}, these include flow in a wavy channel~\cite{maass96direct}, flow over a curved backward facing step~\cite{bentaleb2012large-eddy}, flow in a converging--diverging channel~\cite{laval2011direct}, and flow over periodic hills~\cite{breuer2009flow}. However, even the configurations in such an extensively studied class of flows can differ from each other dramatically in terms of obstacle-height based Reynolds number (ranging from $O(10^2)$ to $O(10^5)$), periodicity, and flow reattachment.

Similar to other application fields of machine learning, training and testing of data-driven turbulence models require data from \emph{systematically} and \emph{continuously} varied flow conditions (e.g., Reynolds number, geometry, angle of attack, and pressure gradient) with a maximum coverage of parameter space. Such a parameter-sweeping database is essential for researchers to test the predictive generality of their data-driven models. 
For example, while each data-driven turbulence model has its own characteristics and generalization capability, one can reasonably expect a data-driven model trained on one class of flows (or a larger dataset consisting these flows as a subset) to be able to predict other flows of the same type. For example, a model trained on attached boundary layers of various pressure gradients should be able to predict a flow with a pressure gradient unseen in the training data, possibly even higher or lower than any of the flows, i.e., involving extrapolation. However, it would be demanding if it is used to predict flows with physics absent in the training flows, e.g., flow with massive separations. 
Unfortunately, despite the two exceptions above, generally speaking one can state that databases tailored for data-driving modeling of turbulent flows are lacking~\cite{kutz17deep}. As a result, developers of data-driven models have relied on their own in-house datasets~\cite{ling2016reynolds,singh16machine,wu2018physics-informed}, making it difficult to compare different methods rigorously due to the different datasets they used.  A comprehensive, publicly available database specifically built for data-driven turbulence modeling is essential for the growing community.

\subsection{Building a parameter-sweeping database for data-driven turbulence modeling}

Data-driven turbulence modeling requires datasets of flows at systematically varied flow conditions. As such databases are meant to be used for training and evaluating data-driven turbulence models, it is not of high priority to achieve high Reynolds numbers that are close to practical flows (although it is certainly desirable if resource permits). Nevertheless, as a benchmark database, the quality and reliability of data are still of paramount importance.  In light of such unique requirements, we aim to build such a database by using carefully performed Direct Numerical Simulations (DNS). 
It is expected that small-scale experiments can be leveraged to achieve the same objectives. 
On the other hand, large eddy simulations would be less ideal as they involve the modeling of sub-grid scale stresses and thus leads to uncertainties that are difficult to quantify.

We choose separated flows as a starting point to build a database tailed for development and evaluation data-driven turbulence models. Turbulent flows separated from a smoothly contoured surface are ubiquitous in natural and engineered systems from rivers with complex bathymetry to airfoils and gas turbines at off-design conditions (e.g., stall and take-off). Such flows are highly complex and depend on Reynolds number, boundary geometry, the presence of mean pressure gradient and acceleration, among others. 
In particular, these flows are characterized by strong non-local, non-equilibrium effects, which clearly violates the equilibrium assumption (between turbulence production and dissipation) made in most eddy viscosity models. Currently, no models performs generally well in flows with equilibrium turbulence (e.g., attached boundary layer) and in those with non-equilibrium turbulence (separated flows).
A dataset of separating flows of parameterized geometries would be valuable additions to the existing benchmark databases for data-driven turbulence modeling.
The design methodology proposed here for generating databases of parameterized flow configurations can be extended to other geometries such as airfoils of parameterized shapes and angles of attack, or wall-mounted objects in flows at various pressure gradients.

The rest of the paper is organized as follows. Section~\ref{sec:method} introduces the design of the cases and the methodology used in the direct numerical simulations for generating the data. Section~\ref{sec:results} presents an example to illustrate the process of constructing and testing a machine learning based model based on this dataset. Furthermore, selected mean flow features and the output are analyzed to shed light on the working of machine learning based flow prediction. Finally, Section~\ref{sec:conclude} concludes the paper.

\section{Methodology}
\label{sec:method}

\begin{figure}[!htb]
\centering
\subfloat[Schematic of baseline geometry]{\includegraphics[width=0.55\textwidth]{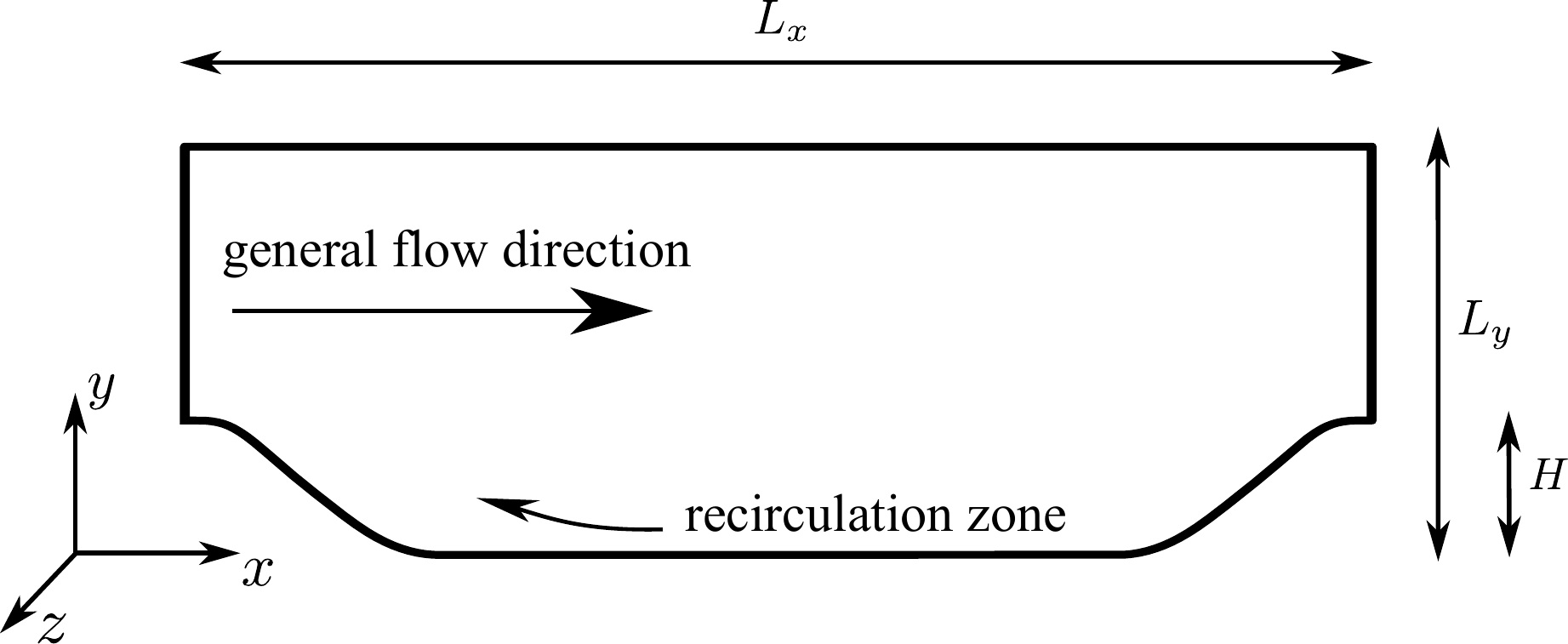}} 
\subfloat[Scheme of slope variation]{\includegraphics[width=0.44\textwidth]{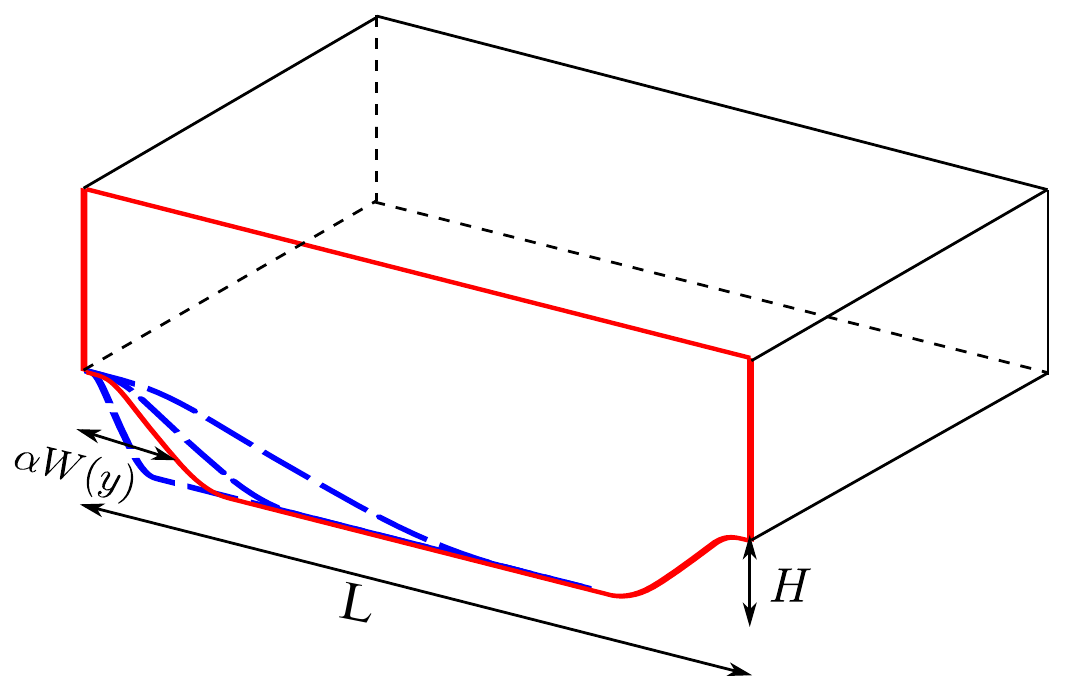}} \\
\subfloat[Geometries on which DNS are performed]{\includegraphics[width=0.99\textwidth]{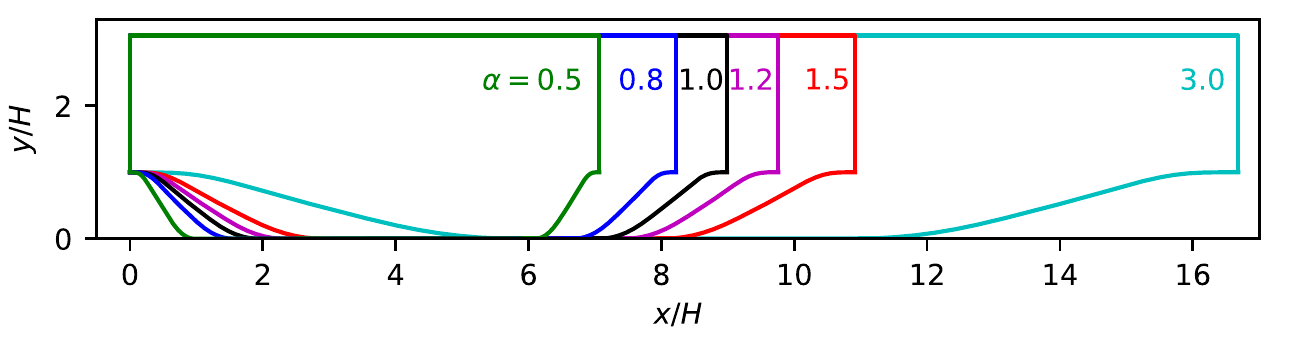}}
\caption{Schematic of the periodic hill geometry and the proposed scheme to achieve different separation patterns by systematically varying the steepness of the hill. 
(a) Computational domain for the flow over the baseline geometry of periodic hills. The $x$, $y$ and $z$ coordinates are aligned with streamwise, wall-normal, and spanwise, respectively. The dimensions of the original geometry are normalized with $H$ with $L_x/H=9$, $L_y/H=3.036$, and $L_z/H=0.1$.
(b) Scheme for slope variation. The solid line denotes the hill profile of the baseline geometry. The dashed lines show different steepness of the profiles, parameterized by multiplying with a factor $\alpha$ to the hill width $W(y)$ of the baseline profile. The length of the bottom flat region is kept constant in this work but can potentially be varied in future efforts.
(c) Geometry variations obtained by scaling the width of the hill by a factor of $\alpha =  0.5$, 0.8, 1.0 (baseline), 1.2, 1.5, and 3.0, for which DNS data are generated. As the length of the flat section is constant, the total horizontal length of the domain is $L_x/H=3.858 \alpha + 5.142$.
\label{fig:schematic}
}
\end{figure}

The flow over periodic hills is widely utilized to evaluate the performance of turbulence models due to the comprehensive experimental and numerical benchmark data at a wide range of Reynolds numbers~\cite{frohlich2005highly,breuer2009flow,rapp2011flow} from $Re=700$ to $10595$. The geometry of the computational domain is shown in Fig.~\ref{fig:schematic}a.  Recently, Gloerfelt and Cinnella performed LES for flows in the same geometry~\cite{gloerfelt2019large} at Reynolds numbers\footnote{The database~\cite{gloerfelt2019benchmark} only contains data up to $Re=19000$ at this time.} up to 37000. However, benchmark data with systematically varied geometry configurations is still lacking. In this work, we performed DNS to build a database of separated flows of the periodic hill geometry~\cite{breuer2009flow} with various steepness ratios (Fig.~\ref{fig:schematic}b, c) at Reynolds number $Re=5600$, which is based on the crest height $H$ and the bulk velocity $U_b$ at the crest. Periodic boundary conditions are applied in the streamwise ($x$) direction, and non-slip boundary conditions are applied at the walls. The spanwise ($z$) direction is homogeneous (periodic boundary conditions for the DNS) and thus the mean flow is two-dimensional.

\begin{figure}[!htb]
\centering
\subfloat[$\alpha = 3$ ]{\includegraphics[width=0.33\textwidth]{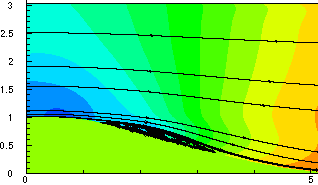}} 
\subfloat[$\alpha = 1.5$]{\includegraphics[width=0.33\textwidth]{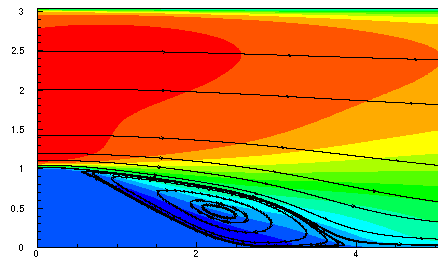}}
\subfloat[$\alpha = 0.5$]{\includegraphics[width=0.33\textwidth]{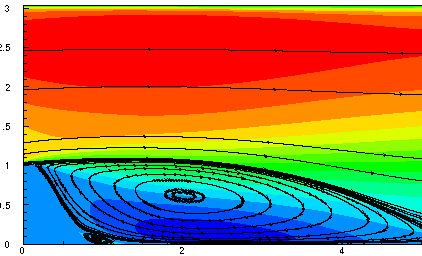}}
\caption{Separated flows come in many sub-regimes depending on the Reynolds number and slope geometry, among others. This figure shows that the size of separation bubble increases as the hill slope becomes steeper in the flow over periodic hills at $Re = 5600$, covering a range from (a) incipient separation to (b) mild separation and (c)massive separation. 
\label{fig:variation}
}
\end{figure}

The variation in geometry (as shown in Fig.~\ref{fig:schematic}c) causes incipiently-separated, mildly-separated, and vastly separated flows to occur (see Fig.~\ref{fig:variation}). The data have been used in our previous works on developing and evaluating data-driven turbulence model based on physics-informed machine learning\cite{wu2019representation,wu2018physics-informed,wu2019reynolds}. Great caution was exercised in ensuring the quality of the simulations by comparing with the previous benchmark data~\cite{breuer2009flow} in both the mean velocities and the Reynolds stresses (detailed in Section~\ref{sec:validate}).

\subsection{Numerical methods}
Direct numerical simulations are performed by solving the forced incompressible Navier-Stokes equations for a fluid with a constant density $\rho$:
\begin{eqnarray}
\frac{\partial \mathbf{u}}{\partial t} &=&
- \frac{1}{\rho}\bnabla p
- \frac{1}{2}\left[\bnabla \left(\mathbf{u} \otimes
\mathbf{u}\right)
+(\mathbf{u} \cdot \bnabla) \mathbf{u} \right]
+ \nu \bnabla^2 \mathbf{u}
+ \mathbf{f}
\label{qdm+f} \\
\boldsymbol{\nabla} \cdot \mathbf{u} &=& 0
\label{divu=0+f}
\end{eqnarray}
where $\otimes$ indicates outer-product of vectors; $p(\mathbf{x},t)$ is the pressure field and $\mathbf{u}(\mathbf{x},t)$ the velocity field; $\mathbf{x}$ and $t$ are spatial and temporal coordinates, respectively. The forcing field $\mathbf{f}(\mathbf{x},t)$ is used to enforce boundary conditions through an immersed boundary method and to ensure a specified mass flux. Note that convective terms are written in the skew-symmetric form, which allows the reduction of aliasing errors while retaining energy conserving for the spatial discretization permitted in the code~\cite{kravchenko&moin97}.

These equations are solved on a Cartesian mesh with the high-order
flow solver \texttt{Incompact3d}, which is based on sixth-order compact schemes
for spatial discretization and a third-order Adams-Bashforth scheme
for time advancement. The code is publicly available on GitHub~\cite{incompact3d-code}. 
To treat the incompressibility condition, a
fractional step method is used, which requires to solve a Poisson equation and it is solved fully in the spectral space. The divergence-free condition is ensured up to machine
accuracy by utilizing the concept of modified
wavenumber. The pressure mesh is staggered from the velocity mesh by half a mesh point to avoid spurious pressure oscillations. The modelling of
the hill geometry is performed with a customized immersed
boundary method based on a direct forcing approach that ensures a
zero-velocity boundary condition at the wall.
Following the strategy of \cite{gautier2014dns}, an artificial flow is reconstructed inside the 2D hill in order to avoid any discontinuities at the wall. More details about the present code and its
validation can be found in \cite{laizet&lamballais09}. The high computational cost of the
present simulations requires the parallel version of this solver. The computational domain is split into
a number of sub-domains (pencils) which are each assigned to an
MPI-process.  The derivatives and interpolations in the $x$-,
$y-$, and $z$- directions are performed in $X$-, $Y$-,
$Z$-pencils, respectively. The 3D FFTs required by the Poisson solver
are also broken down as series of 1D FFTs computed in one direction at
a time. The parallel version of \texttt{Incompact3d} can be used with up to one million computational cores~\cite{laizet&li11}.

The solver \texttt{Incompact3d} has been used recently for a variety of projects ranging from the study of the main features of gravity currents \citep{espath2014two, schuch2018three,
  lucchese2019direct}, performance analysis of active and passive drag
reduction techniques
\citep{mahfoze2017skin,yao2018drag,mahfoze2019reducing}, heat transfer
feature of an impinging jet
\citep{dairay2014turbulent,dairay2015direct}, wake-to-wake interactions in wind farms \citep{deskos2018development,deskos2019turbulence}, and non-equilibrium
scalings of turbulent wakes
\citep{dairay2015non,obligado2016nonequilibrium,zhou2017related}.

\subsection{Flow set-up}

Following the design of the baseline periodic hill geometry, the coordinates of the hill are represented as piecewise third-order polynomial functions (detailed in~\ref{app:hill}). The second hill geometry is described by the same equation with a horizontal translation.

When varying the slope of the hills, we keep its height $H$ constant and change its width. The length of the flat section between the hills, which is 5.142 in the baseline geometry, is kept constant as well. Therefore, total horizontal lengths $L_x$ of the computational domain for the varied geometries are thus given by
$$
L_x/H=3.858 \alpha + 5.142. 
$$
The parameter $\alpha$ changes the width of the hill by elongating the $x$-coordinates
of the geometry. The distance between the first hill geometry and the second hill geometry thus changes with $\alpha$. The reference case corresponds to $\alpha=1$ and a crest-to-crest distance of $L_x = 9$.

\begin{table}
\centering
\caption{Summary of mesh resolution along streamwise- ($x$-), wall-normal ($y$), and spanwise- ($z$-) directions. A stretched mesh in the vertical direction is used with a refinement towards the walls. The height of the first cell along wall-normal direction is $\Delta y_\textrm{wall}/H\approx 6.22 \times 10^{-3}$. The spatial resolution is comparable to those found in the benchmark numerical simulations~\cite{breuer2009flow} at the same Reynolds number $Re = 5600$.}
	\begin{tabular}{P{2cm} P{4cm}}	
		\hline	
		$\alpha$ & mesh ($n_x \times n_y \times n_z$) \\
		\hline
		$0.5$ & $768 \times 385 \times 128$\\
		$0.8$ & $704 \times 385 \times 128$ \\
		$1.0$ & $768 \times 385 \times 128$ \\
		$1.2$ & $832 \times 385 \times 128$ \\
		$1.5$ & $768 \times 385 \times 128$ \\
		$3.0$ & $768 \times 385 \times 128$ \\
		\hline
	\end{tabular}
\label{tab:pehill-mesh-info}
\end{table}

The initial condition for the streamwise velocity field is given by
$$
u(y)=1-\left (\frac{y}{H} \right)^2
$$
while the initial conditions for the vertical and spanwise velocity field are equal to zero.
The mass flow rate for all the simulations is kept constant in time via the imposition of a constant pressure gradient at each time step. The data are collected after a transitional period, from the point when the flow is fully developed. A simulation time step $\Delta t = 0.0005H/U_b$ is used. For all the calculations, turbulent statistical data have been collected over a time period $T = 150H/U_b$.

\begin{figure}[!htb]
\centering
\includegraphics[width=0.99\textwidth]{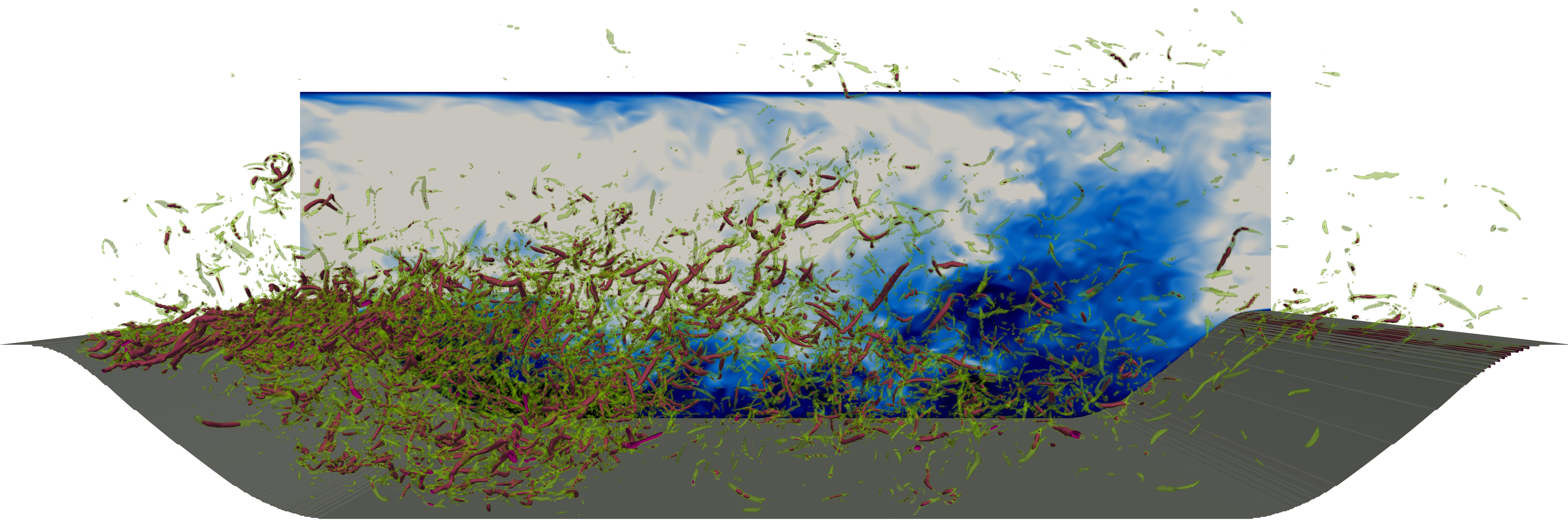}
\caption{Perspective view of vorticity $\Omega$ modulus isosurfaces $|\Omega| = 100U_b/H$ (green/lighter grey) and $|\Omega| = 200U_b/H$ (pink/darker grey) for the simulation with the baseline geometry ($\alpha=1$).
\label{fig:3dview}
}
\end{figure}

An example visualization of the instantaneous flow field is presented in Figure~\ref{fig:3dview} from the simulation with the baseline geometry ($\alpha=1$). One can clearly see the highly three-dimensional nature of the flow structures that develop around the recirculation region of the flow. Intense elongated small-scale flow structures are present when the flow separates at the first hill crest while there is only a small number of intense turbulent structures after the reattachment of the flow.

\subsection{Mesh sensitivity study and validation to benchmark data}
\label{sec:validate}

\begin{figure}[!htb]
\centering
 \hspace{2em}\includegraphics[width=0.56\textwidth]{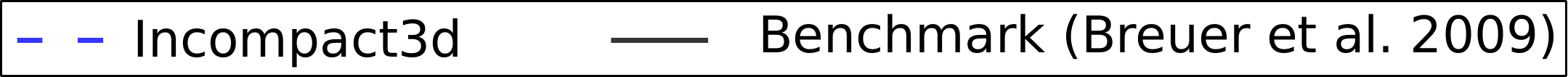} \\
 \subfloat[Mean velocity $U_{x}$]{\includegraphics[width=0.69\textwidth]{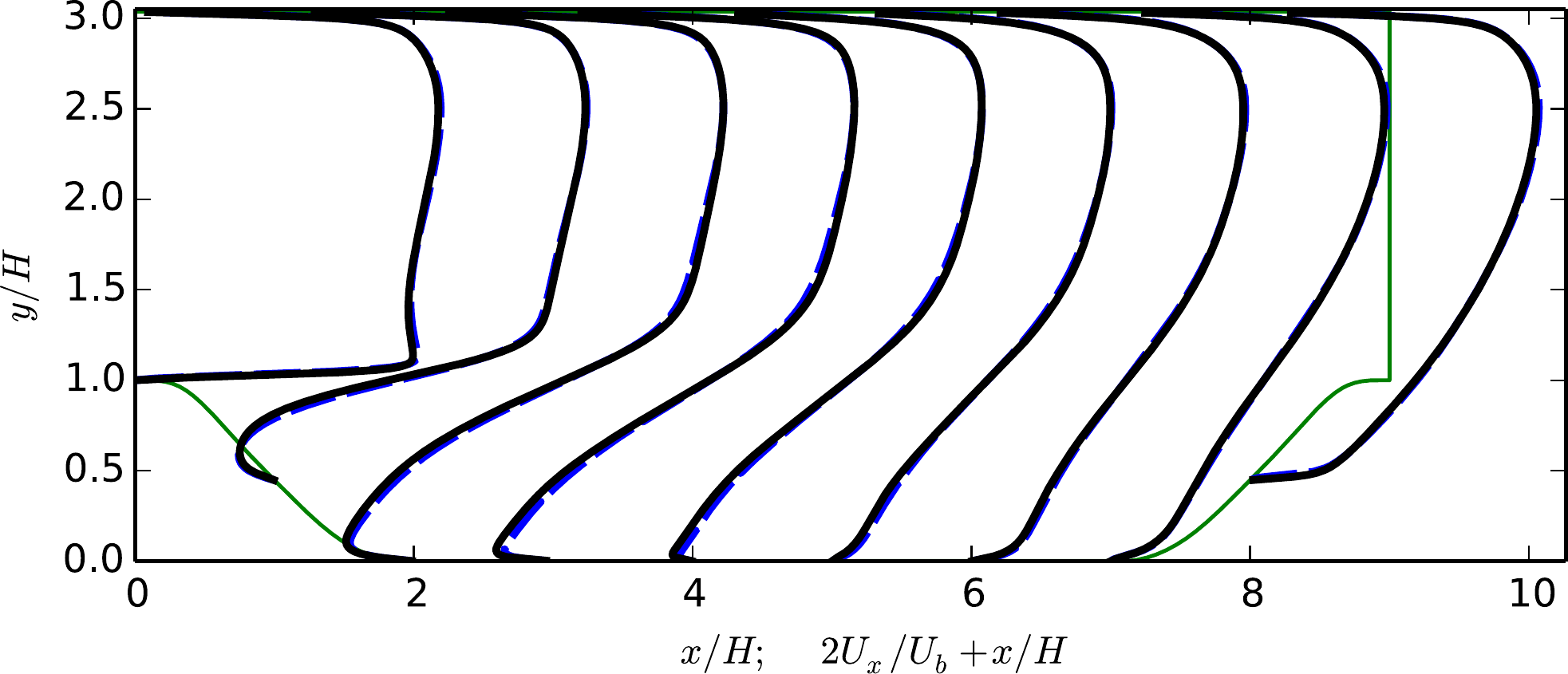}}\\
 \subfloat[Normal stress $\tau_{xx}$]{\includegraphics[width=0.69\textwidth]{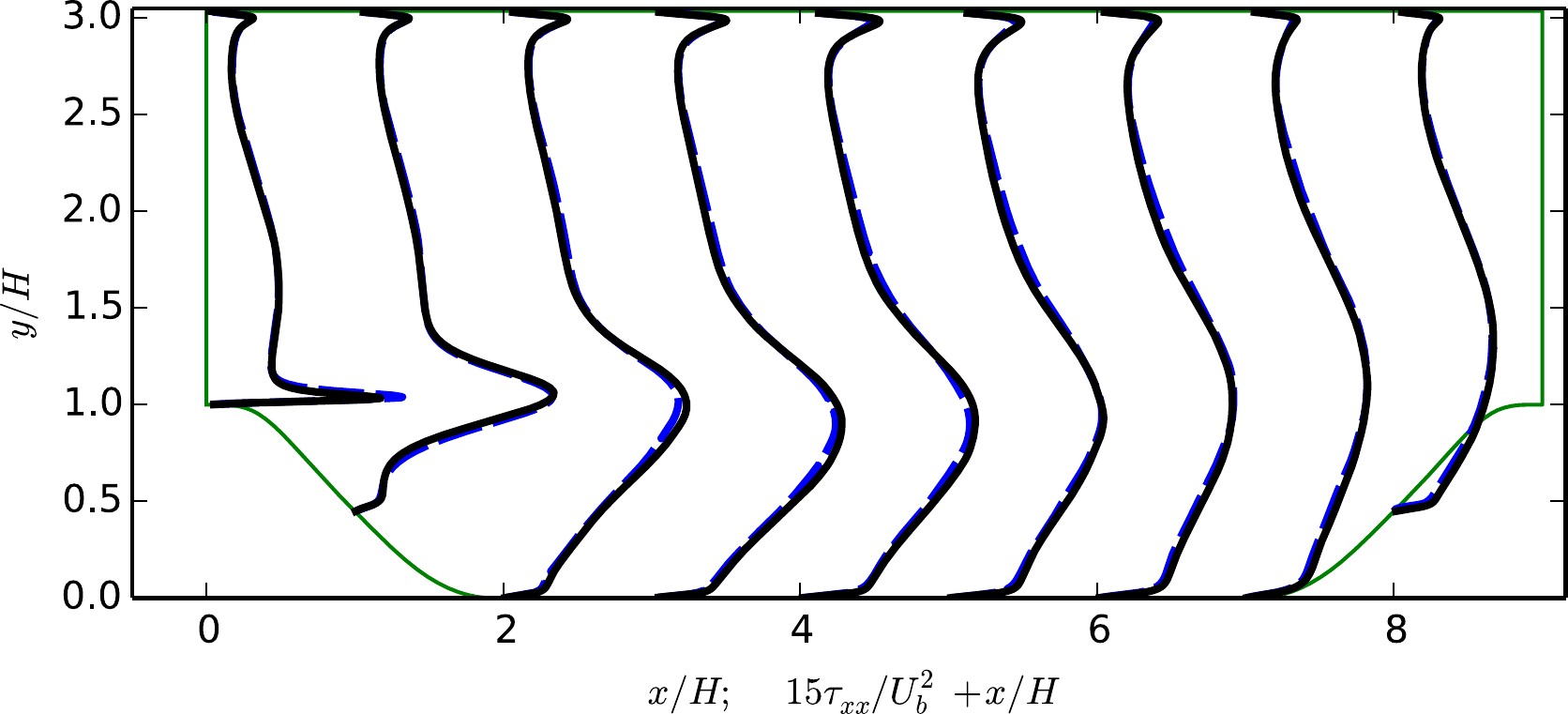}}\\
 \subfloat[Shear stress $\tau_{xy}$]{\includegraphics[width=0.69\textwidth]{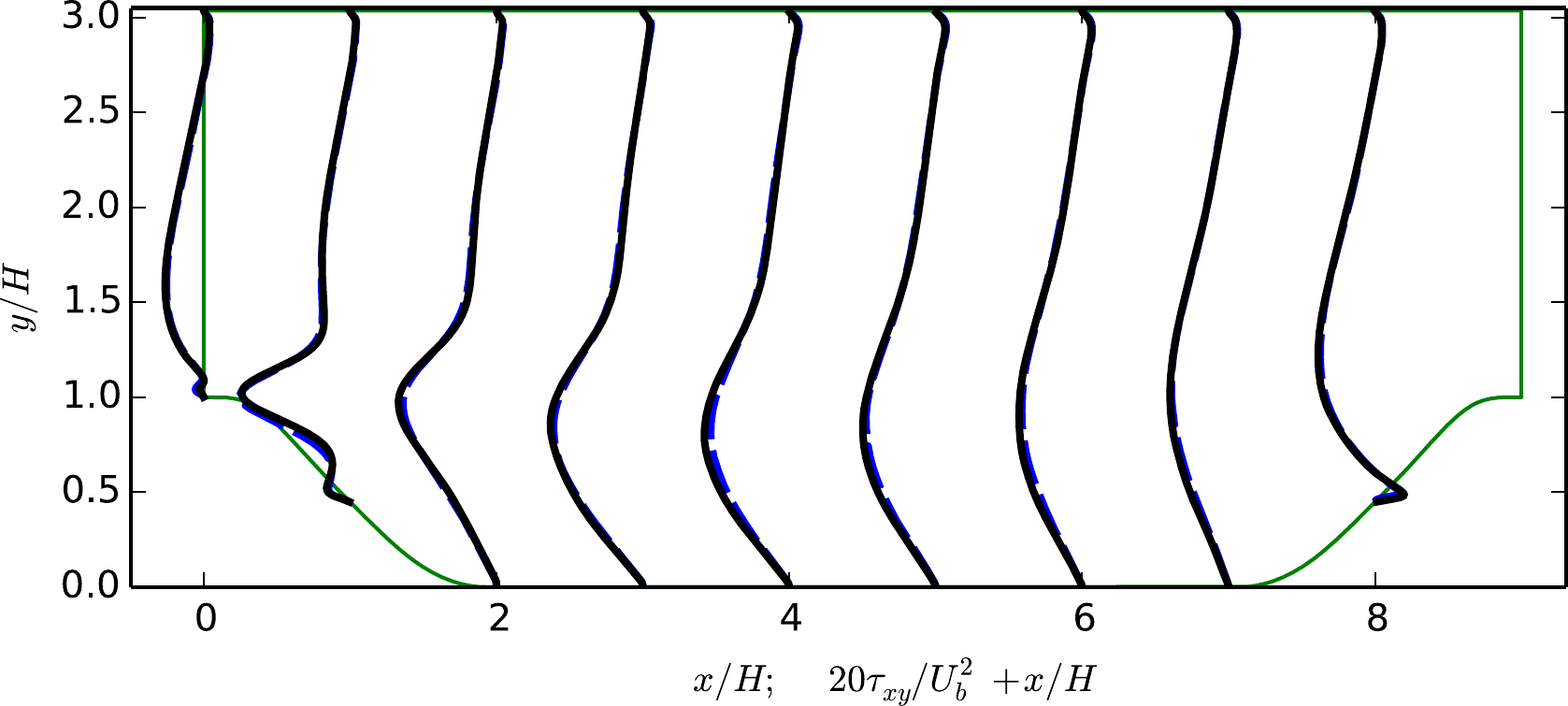}}
\caption{Validation of \texttt{Incompact3d} (dashed lines) against the benchmark data of Breuer et al.~\cite{breuer2009flow} (solid lines), showing profiles of (a) mean velocities $U_x$, (b) turbulent normal stresses $\tau_{xx}$, and (c) turbulent shear stresses $\tau_{xy}$ at nine different streamwise locations $x/H = 0, 1, \ldots, 8$.
\label{fig:validation-Breuer}
}
\end{figure}

We have performed careful validations with existing simulations of flow over periodic hills. Figure~\ref{fig:validation-Breuer} shows a comparison of mean horizontal velocity and Reynolds stress components between the results computed by \texttt{Incompact3d} and those presented by Breuer et al.~\cite{breuer2009flow} at $Re = 5600$. It can be seen that excellent agreements were obtained for both the mean velocity and Reynolds stress components. The comparison of mean vertical velocity $U_y$ shows the similar agreement as the comparison of $U_x$ and is thus omitted here. It should be noted that the normal stress component $\tau_{xx}$ is representative of the strength of turbulent kinetic energy, while the shear stress component $\tau_{xy}$ reflects the strength of the shear flow at downstream of the hill crest. Both these two components from the results computed by \texttt{Incompact3d} show a good agreement with those of Breuer et al.~\cite{breuer2009flow}. The comparison of normal stress component $\tau_{yy}$ also showed a similar good agreement as in Fig.~\ref{fig:validation-Breuer}b and is omitted here for brevity.

\begin{figure}[!htb]
\centering
 \hspace{2em}\includegraphics[width=0.4\textwidth]{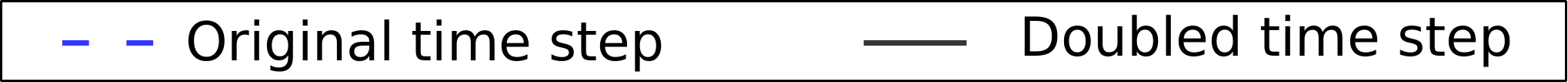} \\
 \subfloat[Normal stress $\tau_{xx}$]{\includegraphics[width=0.49\textwidth]{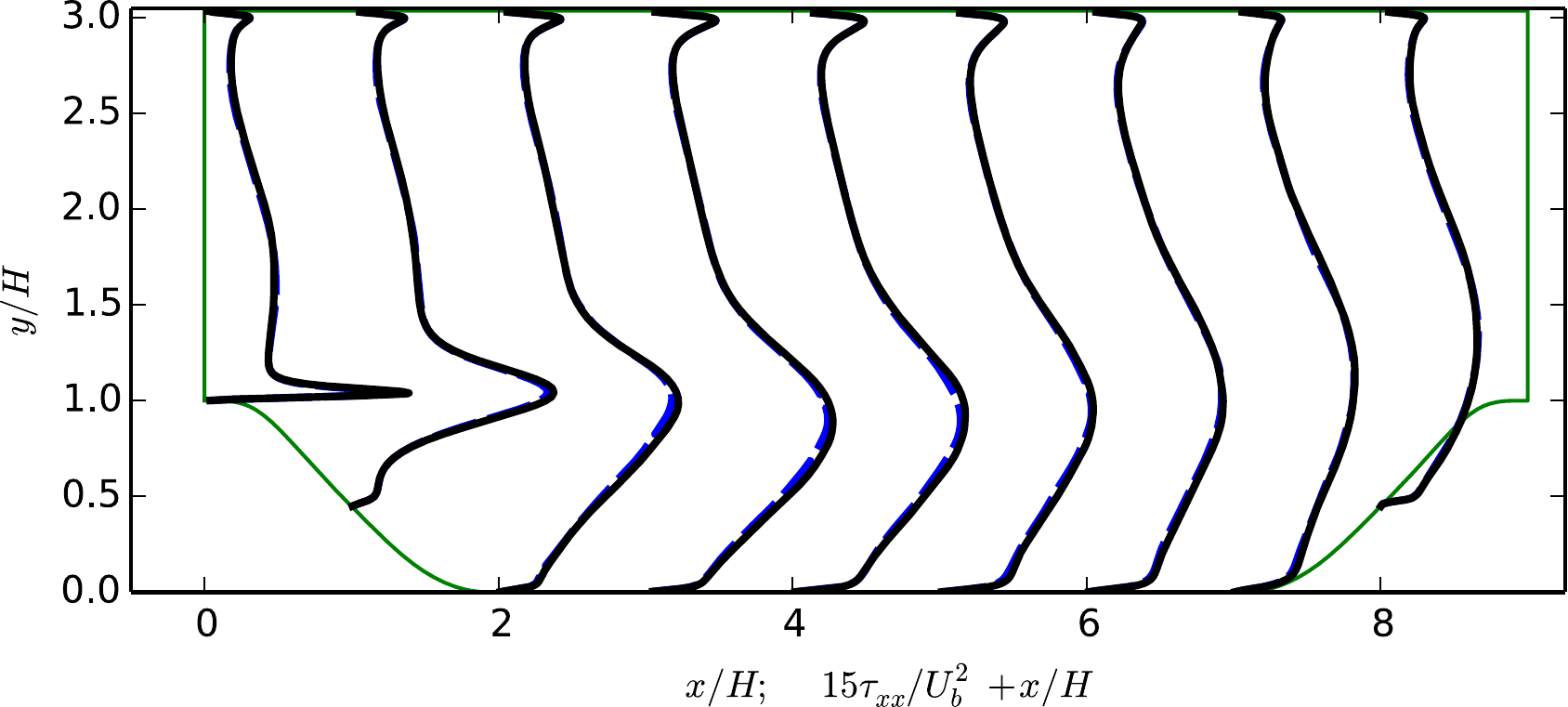}} \hspace{0.1em}
 \subfloat[Shear stress $\tau_{xy}$]{\includegraphics[width=0.49\textwidth]{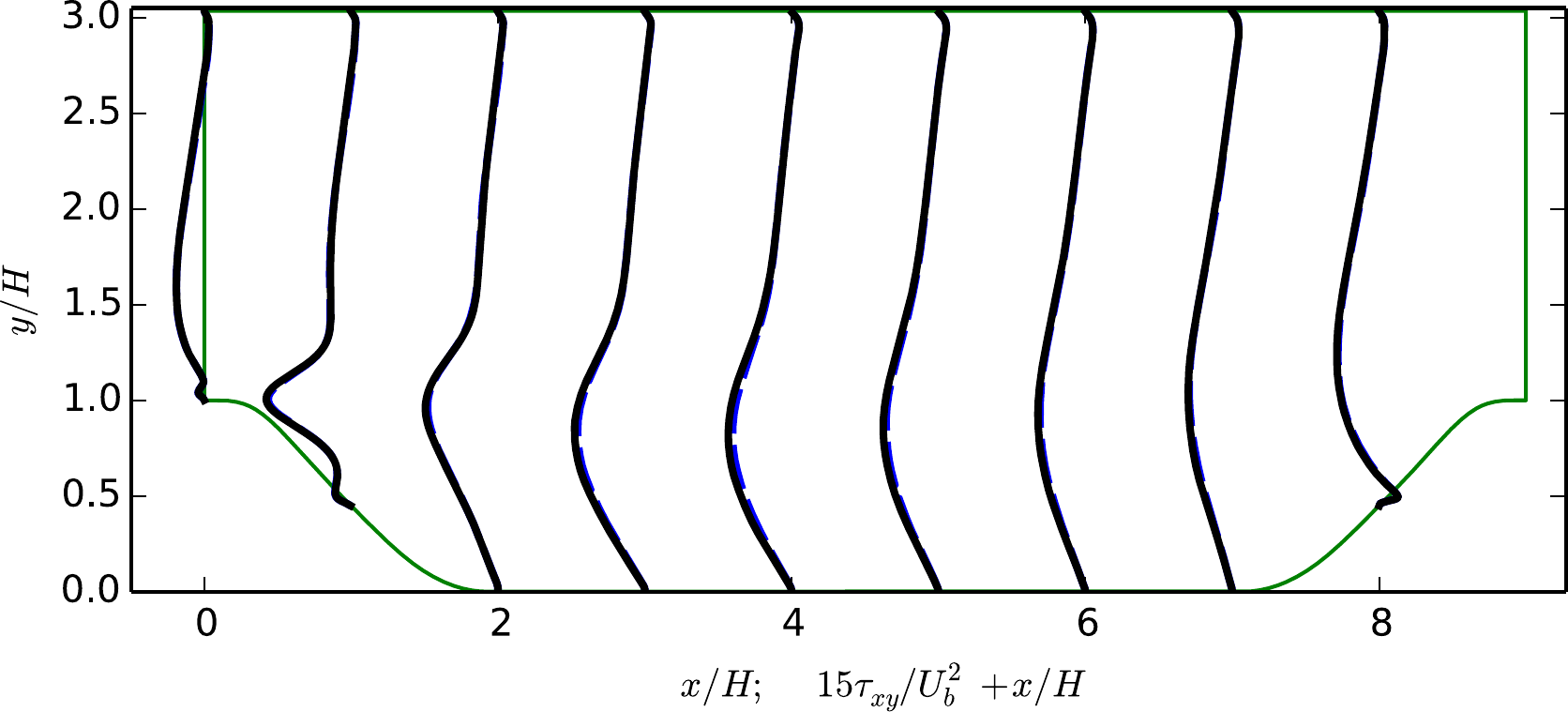}}
\caption{Demonstration of \textbf{time-step size sensitivity} for \texttt{Incompact3d} results with baseline time step $\Delta t$ and doubled time step $\Delta t^\prime=2\Delta t$, showing profiles of (a) turbulent normal stresses $\tau_{xx}$ and (b) turbulent shear stresses $\tau_{xy}$ at nine different streamwise locations $x/H = 0, 1, \ldots, 8$.
\label{fig:validation-T}
}
\end{figure}

\begin{figure}[!htb]
\centering
 \hspace{1em}\includegraphics[width=0.6\textwidth]{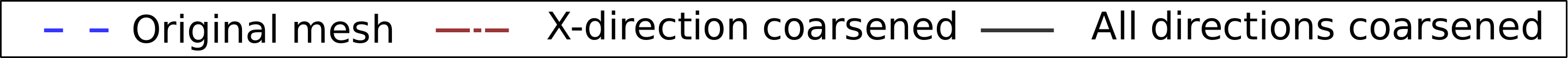} \\
 \subfloat[Normal stress $\tau_{xx}$]{\includegraphics[width=0.49\textwidth]{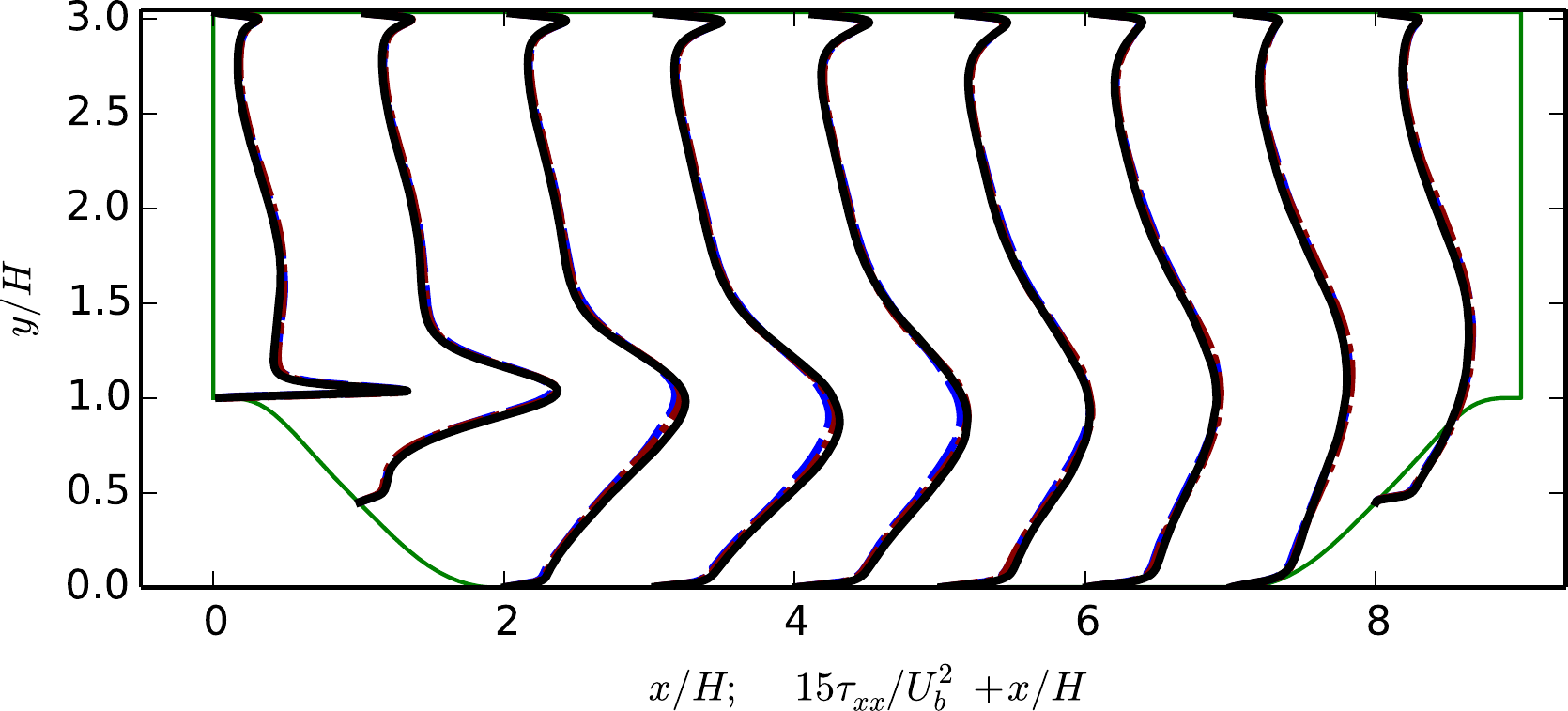}}\hspace{0.1em}
 \subfloat[Shear stress $\tau_{xy}$]{\includegraphics[width=0.49\textwidth]{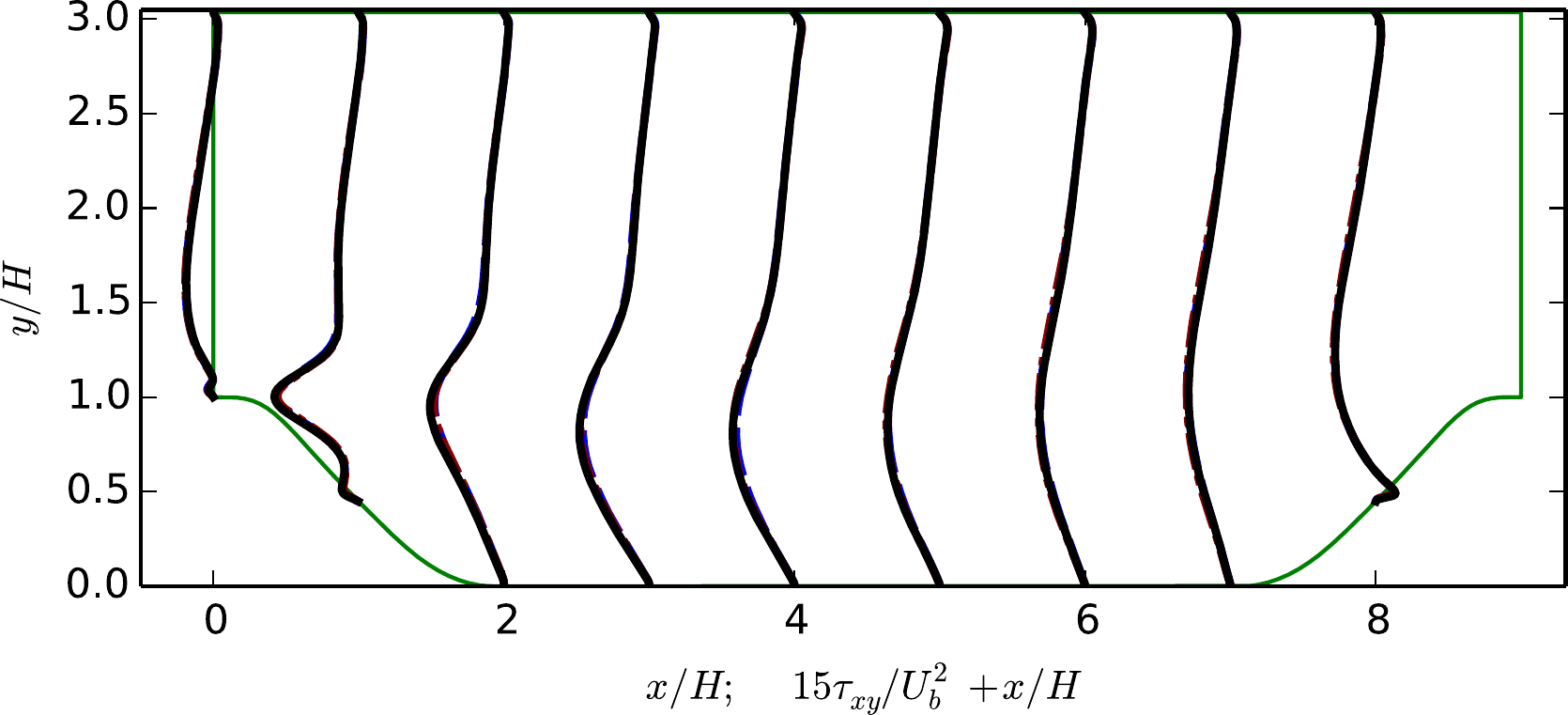}}
\caption{\textbf{Mesh convergence study} on the baseline geometry ($\alpha=1$) for \texttt{Incompact3d} results with original mesh ($n_x\times n_y=768\times 385$), mesh coarsened along x direction ($512\times 385$) and mesh coarsened along both $x$ and $y$ directions ($512\times 257$), showing profiles of (a) turbulent normal stresses $\tau_{xx}$ and (b) turbulent shear stresses $\tau_{xy}$ at nine different streamwise locations $x/H = 0, 1, \ldots, 8$.
\label{fig:validation-mesh}
}
\end{figure}

In order to demonstrate the convergence of resolved turbulence stresses in our database, we further performed simulations with either a larger time step $\Delta t^\prime=2\Delta t$ or coarser spatial resolutions. The results from these simulations are compared with the results of the original simulation in Figs.~\ref{fig:validation-T} and~\ref{fig:validation-mesh}. It can be seen that the turbulence stresses $\tau_{xx}$ and $\tau_{xy}$ do not show any noticible changes by coarsening the temporal or spatial resolutions of the original simulation. Therefore, the convergence of resolved turbulence stresses has been achieved with the resolutions of the simulations presented here.

\section{Data Analysis, Machine Learning, and Interpretation}
\label{sec:results}

The data generated by performing DNS on the designed cases as described in Section~\ref{sec:method} are made available in a public GitHub repository~\cite{xiao2019data} and will also be distributed through the NASA Turbulence Modeling Portal~\cite{nasa-database}. The data include mean pressure field, mean velocities fields, and second order statistics (Reynolds stress fields). As such, our dataset are primarily valuable for developing RANS-based turbulence models. Based on the current literature, most data-driven turbulence models are constructed from these quantities or those derived therefrom, such as strain-rate and rotation-rate (the symmetric and anti-symmetric parts of velocity gradient), pressure gradient, and streamline curvature. We recognize that future models may need high-order statistics (e.g., velocity triple correlation tensor, pressure-strain rate tensor), which will be recorded in future simulations. 

From these mean pressure and velocity fields, various other features can be derived. Table~\ref{tab:features} shows a set of five hand-crafted mean flow features following that of Ling and Templeton~\cite{ling2015evaluation} and Wang et al.~\cite{wang2017physics-informed}. In a typical RANS simulation, these quantities would be available, and such mean flow features have been used to predict the reliability of linear eddy viscosity models~\cite{ling2015evaluation}, the discrepancies of the RANS-modeled Reynolds stresses~\cite{wang2017physics-informed,wang2019prediction}. They can also be used to predict Reynolds stress itself or the eddy viscosity~\cite{zhu2019machine} without relying on any baseline RANS models~\cite{ling2016reynolds}. The specific choice of mean flow features (inputs) and the output quantities (Reynolds stress, eddy viscosity, or their discrepancies, or the symbolic form of the Reynolds stress--strain-rate function) are what differentiate the data-driven models.

\subsection{Problem formulation}

\begin{figure}[!htb]
  \centering 
\includegraphics[width=0.45\textwidth]{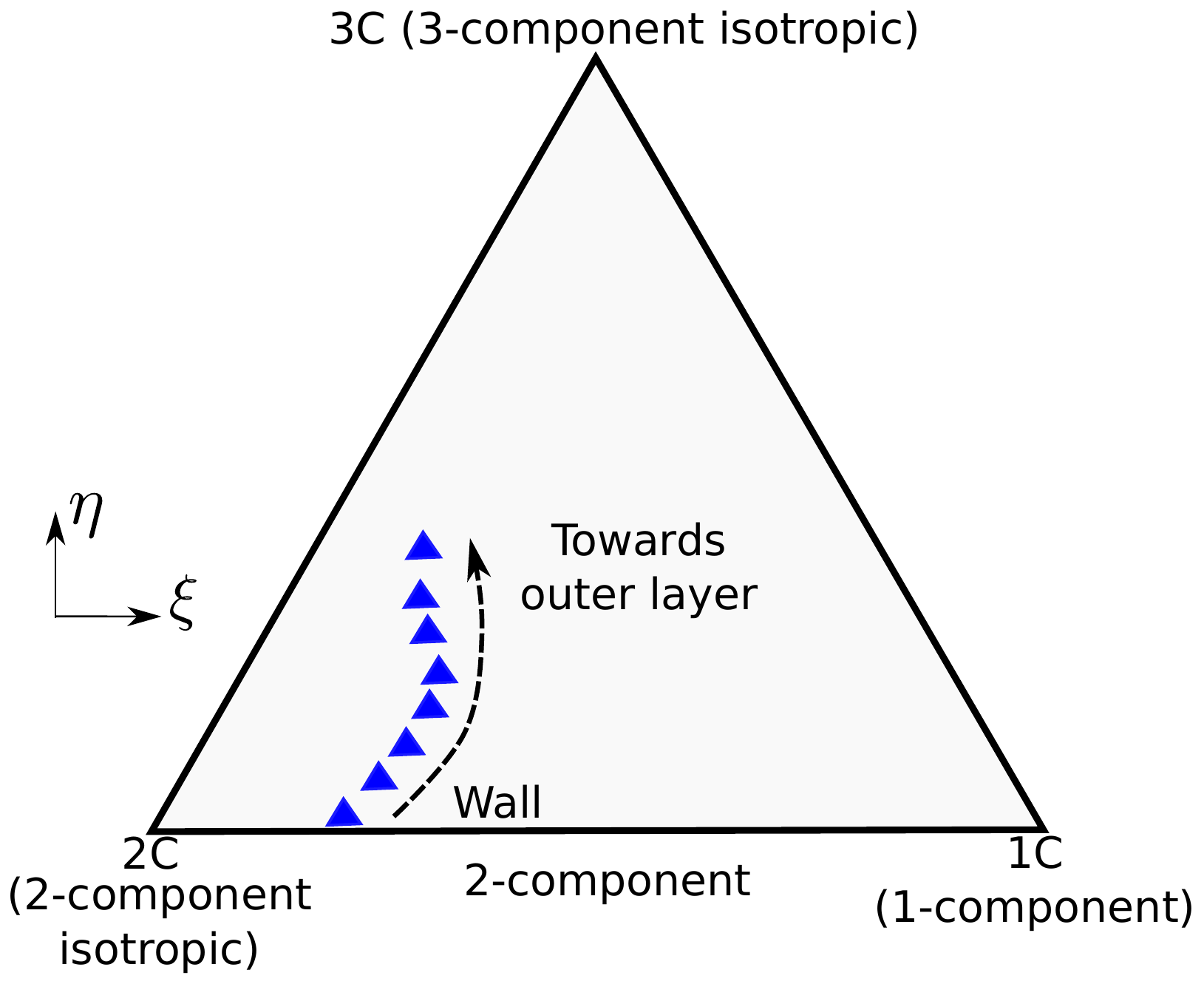}
\caption{The Barycentric triangle encloses physically realizable Reynolds stresses~\cite{banerjee2007presentation}. The position in the Barycentric triangle represents the componentality of the Reynolds stress anisotropy, e.g., three-component isotropic state (vertex 3C), two-component isotropic state (vertex 2C), one-component state (vertex 1C), and a combination thereof (interior). The coordinate system places the triangle in the $\xi$-$\eta$ plane with the origin $(0, 0)$ located at 2C.}
  \label{fig:bary-scheme}
\end{figure}

We aim to present a methodology of generating DNS benchmark datasets of systematically varying flow configures for data-driven turbulence modeling. As such, we refrain from advocating any particular data-driven turbulence model. 
Here we use an example to illustrate the possible usage of our dataset for training and testing data-driven models. 
To this end, machine learning models are constructed to predict Reynolds stress anisotropy based on the mean flow features $\mathbf{q}$ shown in Table~\ref{tab:features}, which is a subset of what were used in~\cite{wang2017physics-informed}. The original features that are based on quantities from RANS solvers (e.g., turbulent kinetic energy $k$ and dissipation rate $\varepsilon$) are omitted here as no RANS simulations are performed in this work. Only DNS data are used in this example. The output quantities as the anisotropy of the Reynolds stress tensor defined based on the following eigen-decomposition:
\begin{equation}
  \label{eq:tau-decomp}
  \boldsymbol{\tau}
  = 2 k \left( \frac{1}{3} \mathbf{I} + \mathbf{V} \Lambda \mathbf{V}^T \right)
\end{equation} 	
where $k$ is the turbulent kinetic energy; $\mathbf{I}$
is the second order identity tensor; $\mathbf{V}$ and
$\Lambda = \textrm{diag}[\lambda_1, \lambda_2, \lambda_3]$
are the eigenvectors and eigenvalues of the Reynolds stress anisotropy tensor.
The eigenvalues are further mapped to
the Barycentric coordinates $(C_1,C_2, C_3)$ as follows~\cite{emory2011modeling}:
	    \begin{equation}
          C_1  = \lambda_1 - \lambda_2,  \quad
          C_2  = 2(\lambda_2 - \lambda_3), \quad \text{and} \quad
          C_3  = 3\lambda_3 + 1 .
          \label{eq:lambda2c}
        \end{equation}  
The coordinate of a point $\boldsymbol{\xi} \equiv (\xi, \eta)$ in the Barycentric triangle can be expressed as that of the three vertices ($\boldsymbol{\xi}_{1c}$, $\boldsymbol{\xi}_{2c}$, and $\boldsymbol{\xi}_{3c}$), i.e.,
$\boldsymbol{\xi} = 	
C_1 \boldsymbol{\xi}_{1c} + C_2 \boldsymbol{\xi}_{2c} +
C_3\boldsymbol{\xi}_{3c}$. The Barycentric triangle and the $\xi$-$\eta$ coordinate system are illustrated in Fig.~\ref{fig:bary-scheme}.
Our example problem thus consists of
(1) learning the functional mapping $\mathbf{q} \mapsto \boldsymbol{\xi}$ from mean flow features $\mathbf{q}$ to a frame-independent quantity $\bm{\xi}$ of the Reynolds stress from training data and (2) assessing the predictive performance of the learned function on a new flow.

  \begin{table}[htbp] 
  \centering
  \caption{
    Non-dimensional flow features used as input in the example machine-learning models.  The normalized feature $q_\beta$ is
    obtained by normalizing the corresponding raw features value $\hat{q}_\beta$ with normalization
    factor $q^*_\beta$ according to $q_\beta = \hat{q}_\beta / (|\hat{q}_\beta| + |q^*_\beta|)$. Repeated indices $i$, $j$, $k$, $l$, imply summation. Notations are as follows: $U_i$ is mean velocity,  
    $\mathbf{S}$ is the strain rate tensor,
    $\boldsymbol{\Omega}$ is the rotation rate tensor, $\boldsymbol{\Gamma}$ is unit tangential velocity vector,  $D$ denotes material derivative,
    and $L_c$ is the characteristic length scale of the mean flow. $\| \cdot \|$ and 
    $|\cdot|$ indicate matrix and vector norms, respectively.  }
\label{tab:features}
\begin{tabular}{P{2.0cm} | P{5.0cm}  P{3.0cm}  P{4.0cm} }	
  \hline
  feature ($q_\beta$)  & description & raw feature ($\hat{q}_\beta$) &
  normalization factor ($q^*_\beta$)  \\ 
  \hline
  $q_1$  & ratio of excess rotation rate to strain rate (Q-criterion) &
  $\frac{1}{2}(\|\boldsymbol{\Omega}\|^2 - \|\mathbf{S}\|^2)$ & 
  $\|\mathbf{S}\|^2$\\  
  \hline
  $q_4$  & pressure gradient along streamline & $U_k\dfrac{\partial P}{\partial
    x_k}$ & $\sqrt{\dfrac{\partial P}{\partial x_j} \dfrac{\partial P}{\partial
      x_j}U_iU_i}$ \\ 
  \hline
  $q_6$  & ratio of pressure normal stresses to shear stresses &
  $\sqrt{\dfrac{\partial P}{\partial x_i}\dfrac{\partial P}{\partial x_i}}$ &
  $\dfrac{1}{2} \rho\dfrac{\partial U_k^2}{\partial x_k}$\\ 
  \hline
  $q_7$  & non-orthogonality between velocity and its gradient & $\left| U_i U_j\dfrac{\partial
      U_i}{\partial x_j} \right|$ &{$\sqrt{U_l U_l \, U_i \dfrac{\partial U_i}{\partial
        x_j}U_k \dfrac{\partial U_k}{\partial x_j}}$ } \\ %
  \hline
  $q_{10}$  & streamline curvature & {$\left|\frac{D
        \boldsymbol{\Gamma}}{ D s}\right|$ where $\boldsymbol{\Gamma} \equiv
    \mathbf{U}/|\mathbf{U}|$, $D s = |\mathbf{U}| Dt$ } & { $\dfrac{1}{L_c}$ }\\
  \hline						 								
\end{tabular}
\end{table}

\subsection{Machine learning models used in this work}
We use two machine learning models, random forests~\cite{breiman2001random} and fully-connected neural networks, to build such mappings and to compare their performances in predictions. A brief introduction of the two models are given below for readers who are not familiar with machine learning.

\begin{figure}[!htb]
  \centering 
\includegraphics[width=0.4\textwidth]{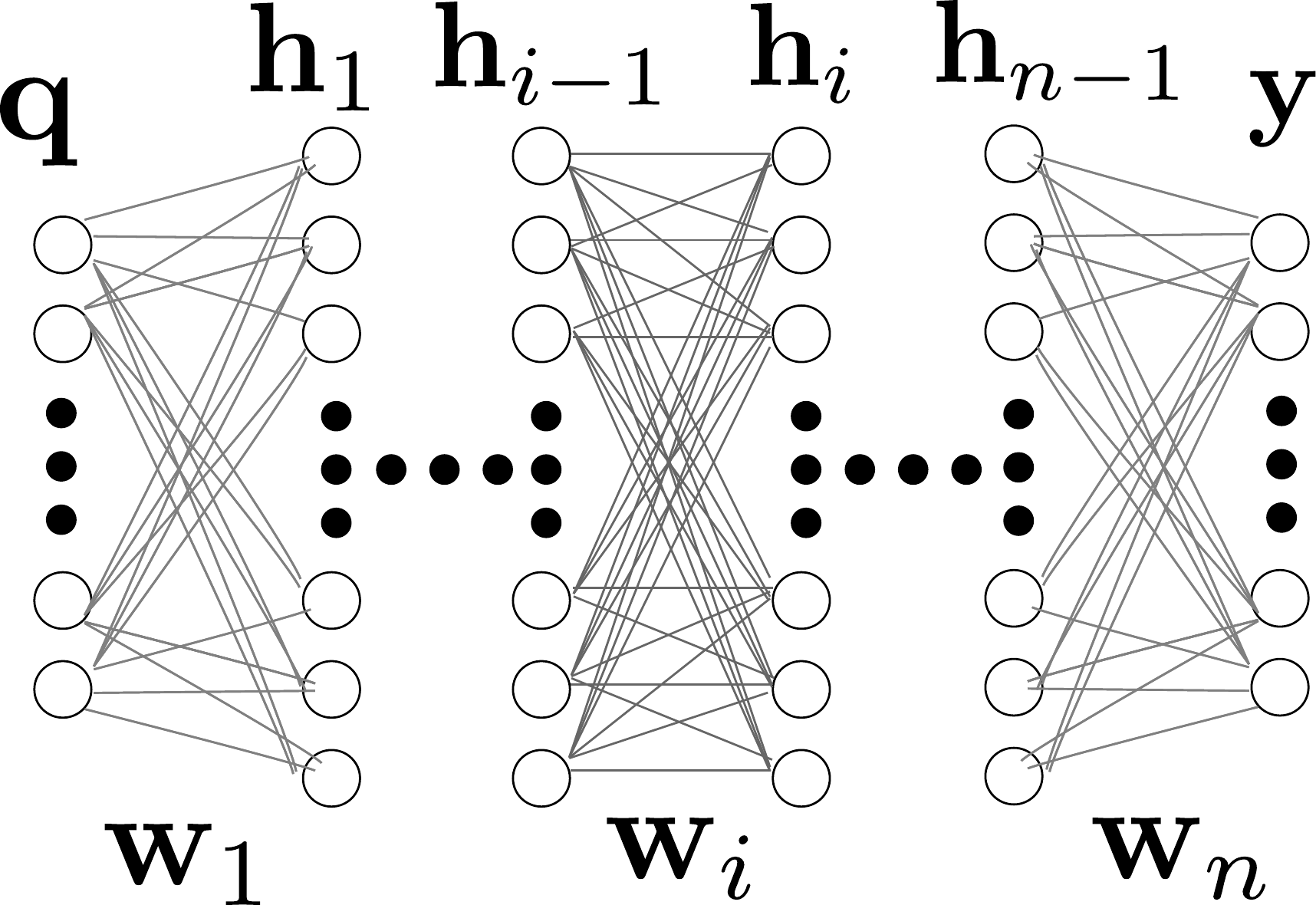}
\caption{Architecture of the fully-connected neural network with $n$ layers (of weights) with $n=7$ used in this work. Each hidden layer has 36 neurons.}
  \label{fig:neural-nets}
\end{figure}

Neural networks are nonlinear functions parameterized by weights $W$ (and biases) that can be learned from data.
They are built by consecutive composition of linear functions (matrix-vector multiplication) followed by nonlinear activation functions.
A neural network in its simplest form is the linear model $\mathbf{y} = \mathbf{W} \mathbf{q}$, which is controlled by weight matrix $\mathbf{W}$ and maps input vector $\mathbf{q}$ to output vector $\mathbf{y}$.  However, such a simple model may lack the flexibility to represent complex functions.  This difficulty can be addressed by introducing one or more intermediate vectors. For example, $\mathbf{h} = \sigma \left(\mathbf{W}^{(1)} \mathbf{q} \right)$ and $\mathbf{y} = \mathbf{W}^{(2)} \mathbf{h}$, or written as composite function $\mathbf{y} = \mathbf{W}^{(2)} \, {\sigma} \left(\mathbf{W}^{(1)} \mathbf{q}\right)$, where $\sigma$ is an activation function such as sigmoid function $\sigma(x) = 1/(1+e^{-x})$ or rectifier linear unit (ReLU), $\sigma(x) = \max(0, x)$.  The functional mapping can be represented by the network diagram in Fig.~\ref{fig:neural-nets}, with each layer corresponding to a vector (e.g., $\mathbf{q}$ (input layer), $\mathbf{h}$ (hidden layer), or $\mathbf{y}$ (output layer)) and each neuron an element therein.  Deep learning utilizes neural networks with many layers.  Neural networks with at least one hidden layer are universal approximators, i.e., they can represent any continuous function on a compact domain to arbitrary accuracy, given enough hidden neurons~\cite{hornik89multilayer}. %
In this work, the input and output layers have 5 and 2 neurons, respectively, which are dictated by the given problem. We use seven hidden layers with 36 neurons in each layer. The ReLu activation function is used. In the training we used 1000 epochs with a learning rate of $0.01$.

Random forests are an ensemble learning method based on decision-tree regression models. Decision-tree models stratify the feature space into different non-overlapping regions so as to minimize the in-region variance of the training data. Then they predict the response for a given test input using the mean of the training observations in the region to which that test input belongs. Regression tree models are simple and useful for interpretation, but their prediction accuracy is generally not competitive compared to, e.g., neural networks.
Random forests improve upon simple regression trees by building an ensemble of trees with bootstrap samples (i.e., sampling
with replacement) drawn from the training data~\cite{friedman2001elements}.  Moreover, when
building each tree, it utilizes only a subset of randomly chosen features to reduces the correlation among the trees in the ensemble and thus decreases the
bias of the ensemble prediction. Random forests have much lower computational costs compared to neural networks and are most robust with only two tuning parameters, i.e.,
the number $N_\text{rf}$ of trees in the ensemble and the number $M$ of selected features. In this work we used an ensemble of $N_\text{rf} = 300$ trees and a subset of features (i.e., $M = 3$) in each stratification (splitting) of the feature space.

\subsection{Predictive performance of machine learning models}
We consider two cases (training/prediction scenarios): 
\begin{enumerate}[label=({\alph*})]
    \item Case 1 involves an interpolation in flow configuration. The training set consists of data from four geometries: $\alpha = \{0.5, 0.8, 1.2, 1.5\}$ and the trained model is used to predict for the flow at $\alpha = 1.0$ (the baseline geometry).
    \item Case 2 involves an extrapolation in flow configuration. The training set consists of $\alpha = \{0.5, 0.8, 1.0, 1.2 \}$, while the flow at $\alpha = 1.5$ is reserved for testing. 
\end{enumerate}
This is summarized in Table~\ref{tab:cases}. In both cases, random forest and neural networks are trained on identical datasets for a fair comparison.

\begin{figure}[htbp]
    \centering
    \includegraphics[width=0.45\textwidth]{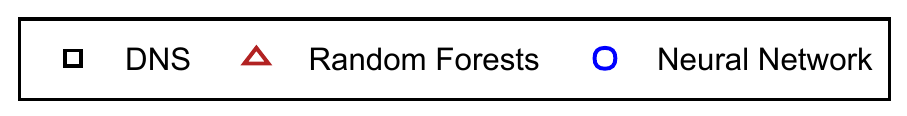}\\
    \subfloat[$x/H = 2$, case 1]{\includegraphics[width=0.33\textwidth]{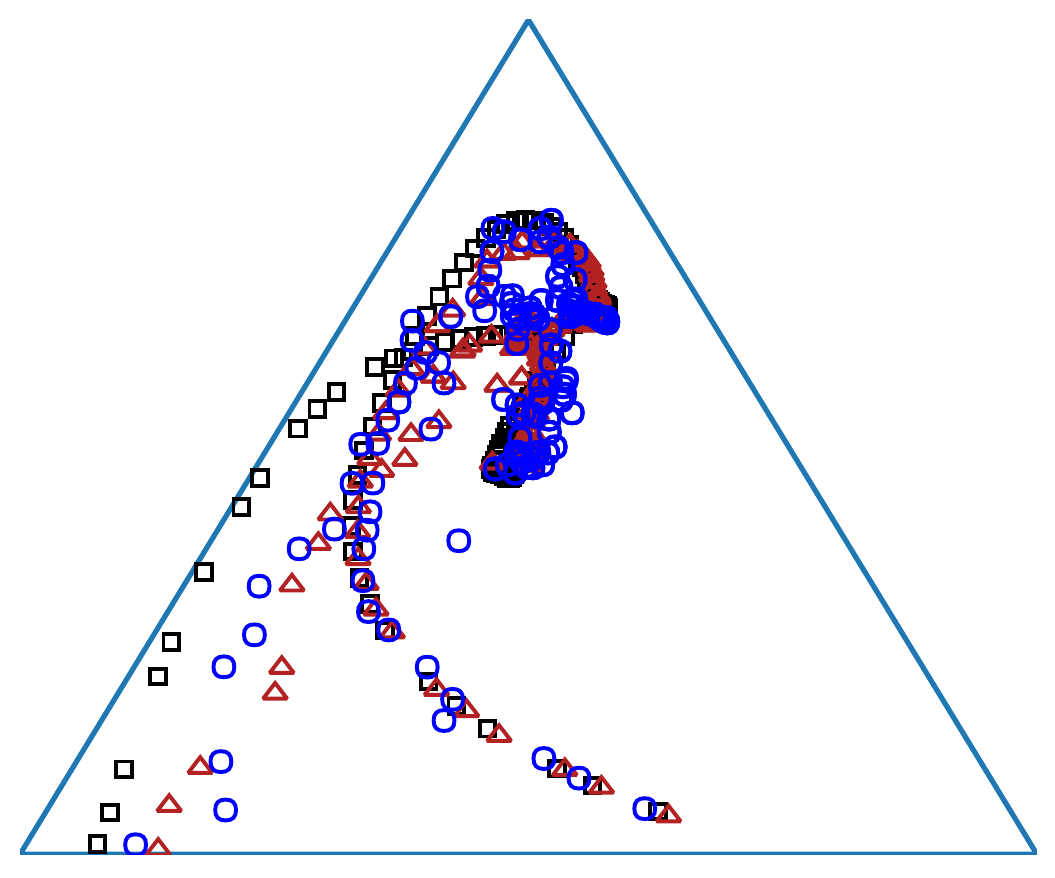}}
    \subfloat[$x/H = 5$, case 1]{\includegraphics[width=0.33\textwidth]{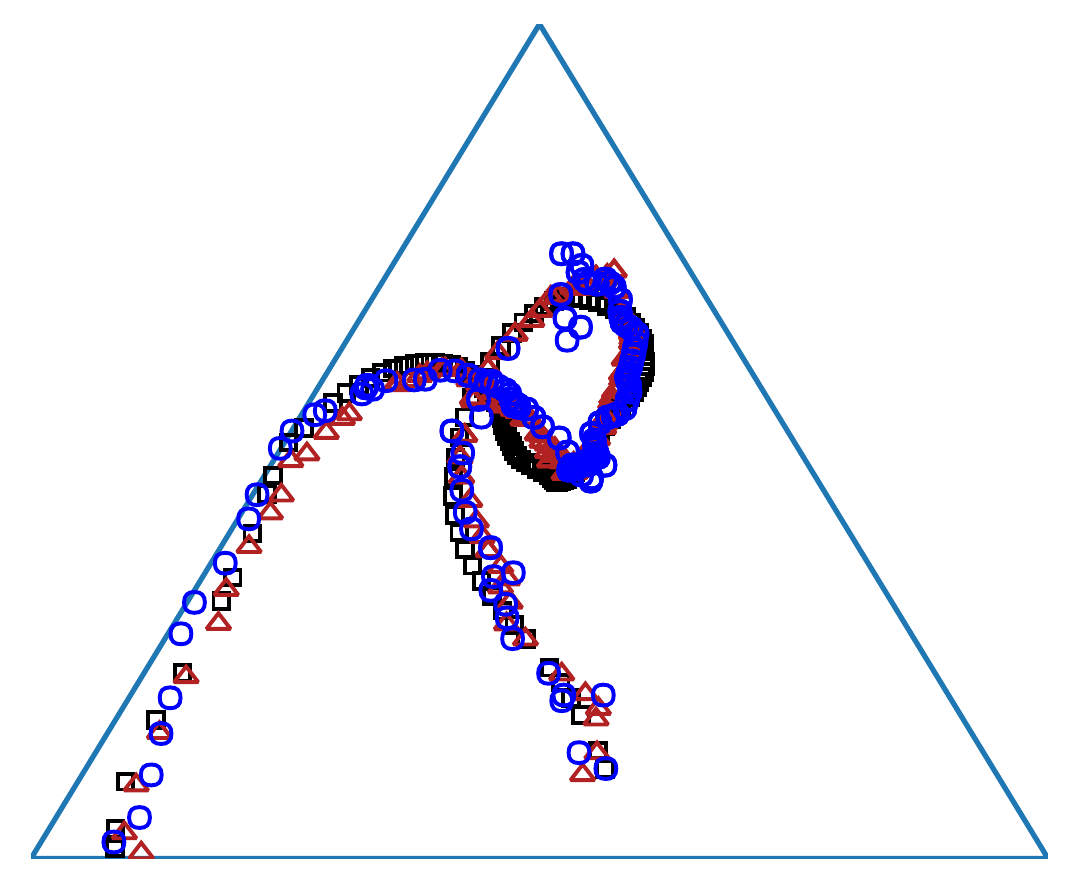}}
    \subfloat[$x/H = 7$, case 1]{\includegraphics[width=0.33\textwidth]{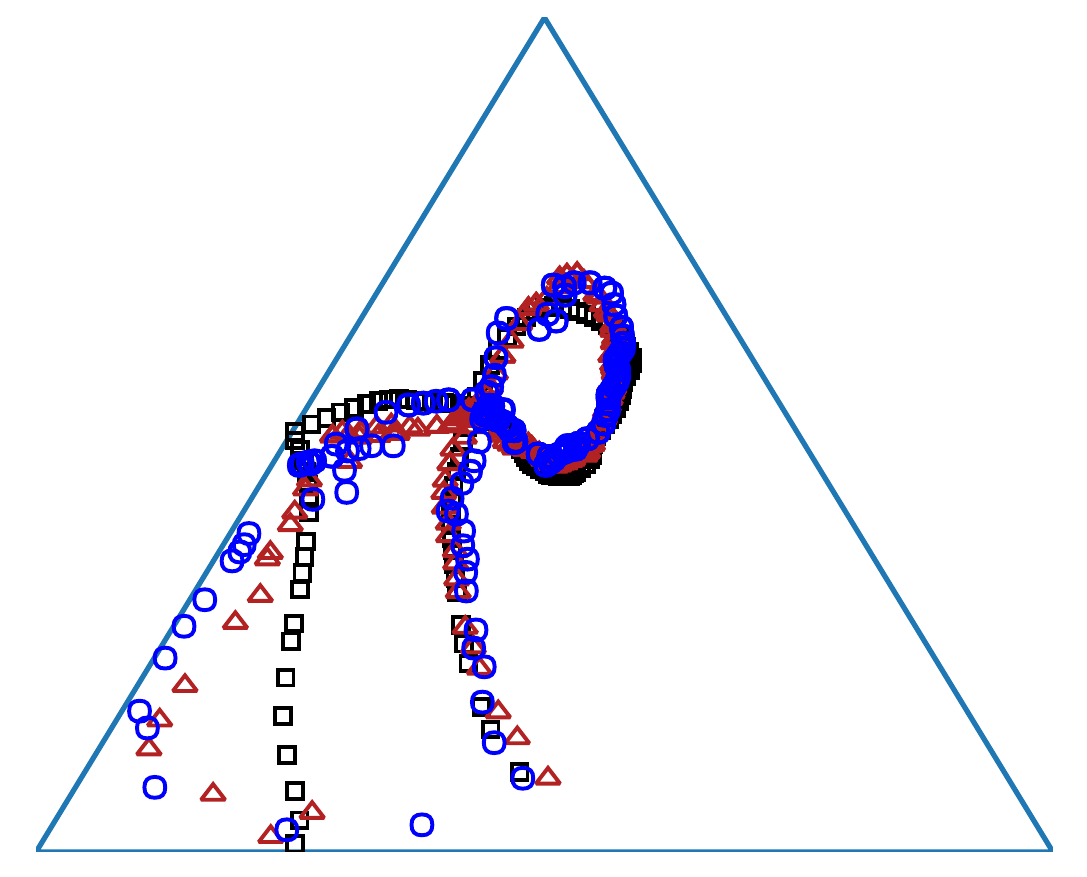}} \\
    \subfloat[$x/H = 2$, case 2]{\includegraphics[width=0.33\textwidth]{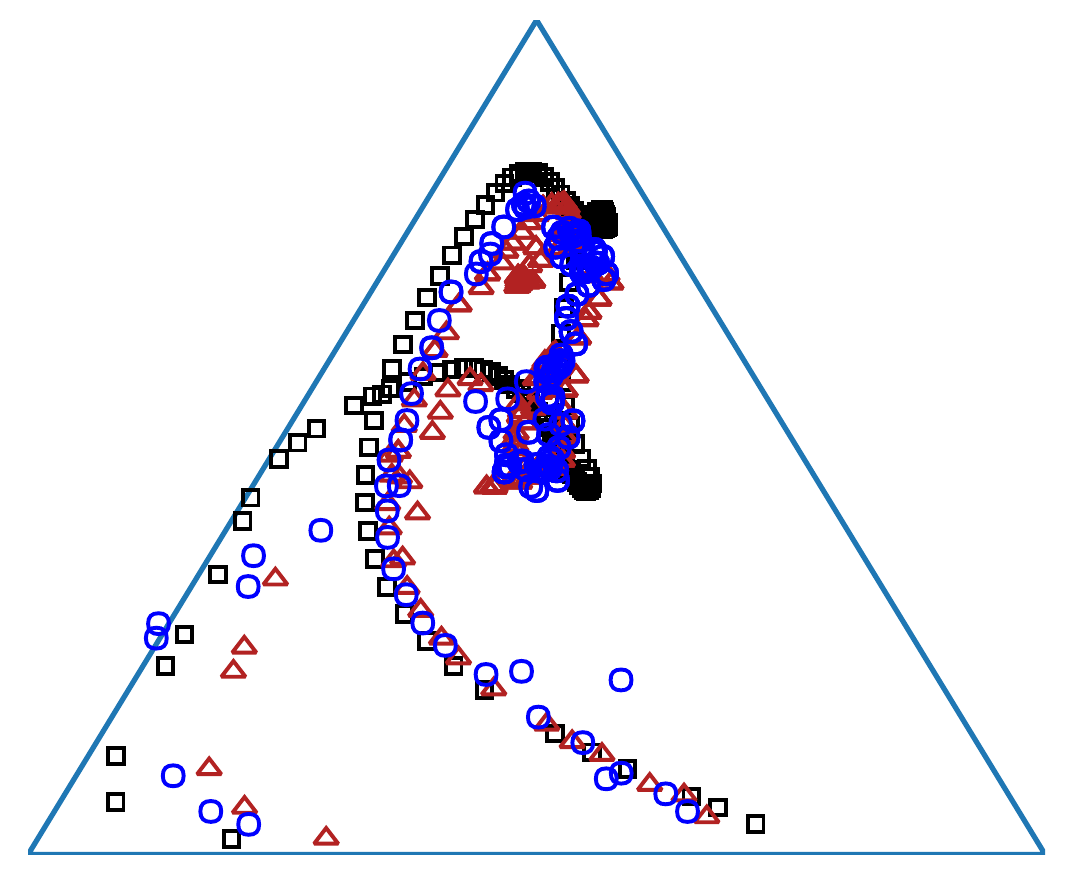}}
    \subfloat[$x/H = 5$, case 2]{\includegraphics[width=0.33\textwidth]{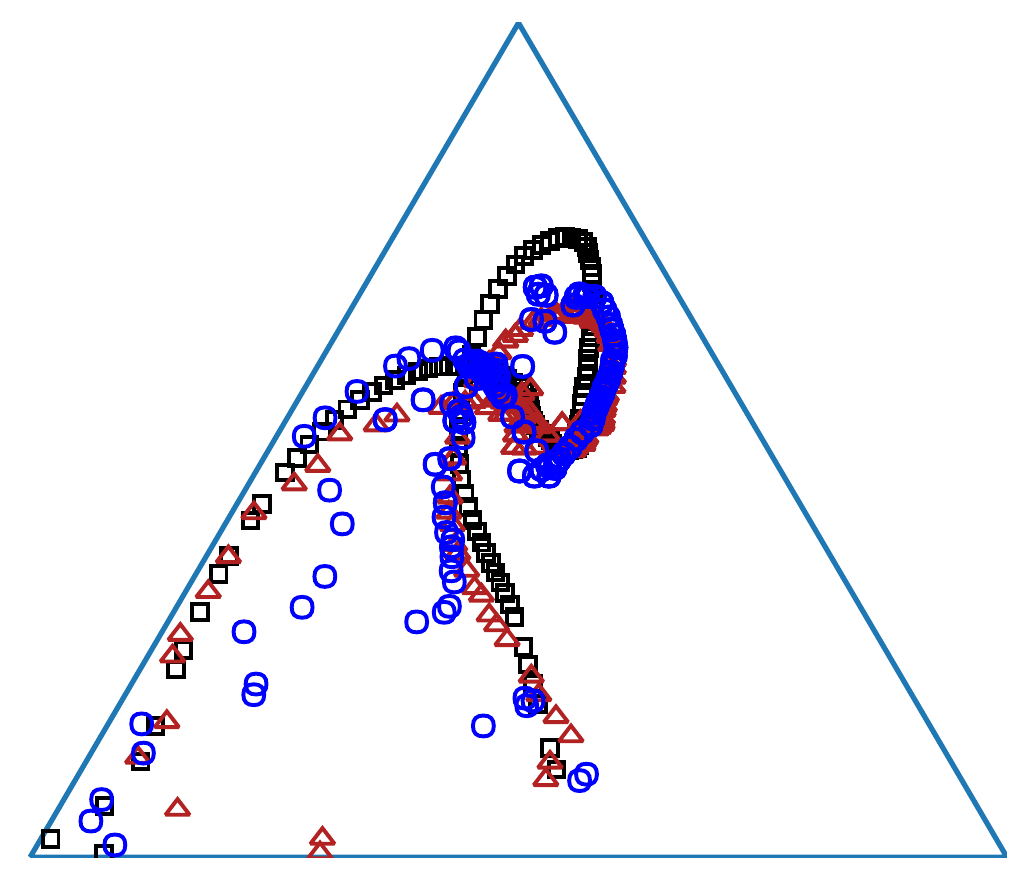}}
    \subfloat[$x/H = 7$, case 2]{\includegraphics[width=0.33\textwidth]{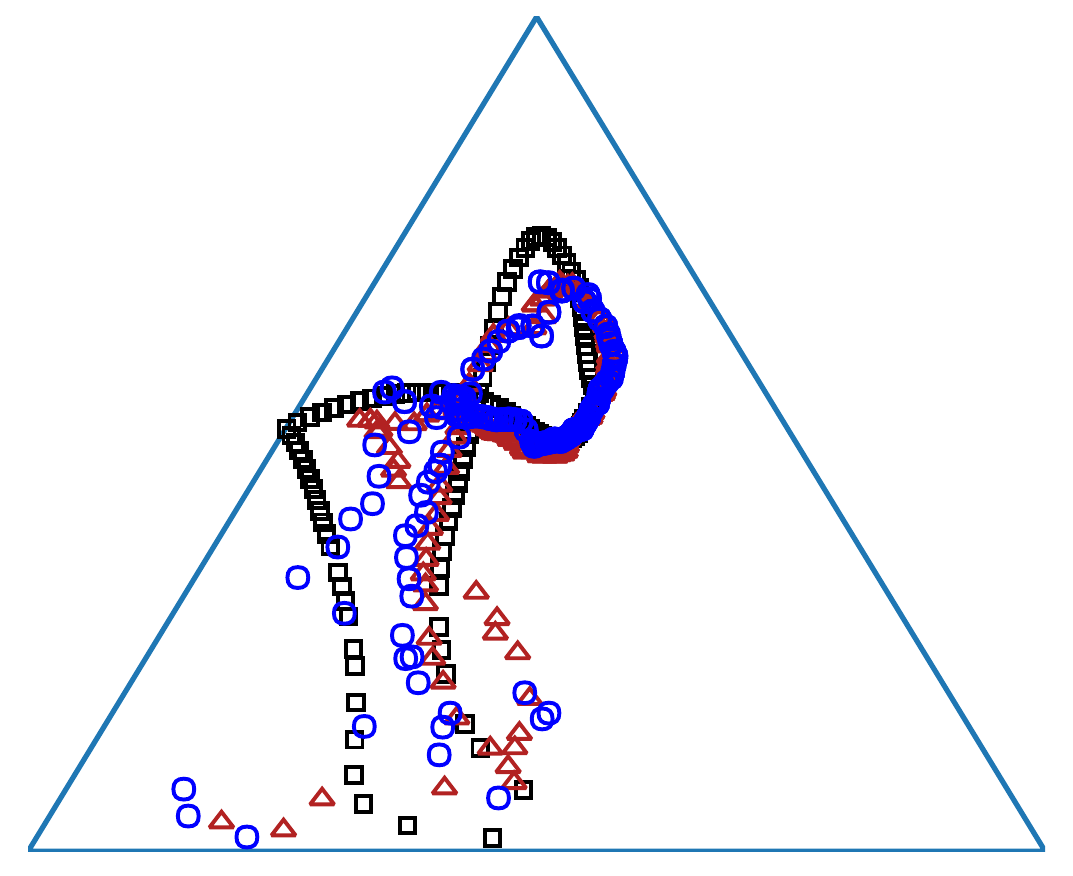}}
    \caption{Predicted Reynolds stress anisotropy for case 1 (interpolation; top panels a--c; the flow with baseline geometry $\alpha=1$ is predicted) and case 2 (extrapolation; bottom panels d--f; the flow with geometry $\alpha=1.5$ is predicted), displayed in Barycentric triangles. Predictions from random forests (RF) and neural networks (NN) are compared with the ground truth (DNS).}
    \label{fig:barycentric_triangles}
    \end{figure}

The predicted anisotropy as represented by the Barycentric plot trajectories  are presented in Fig.~\ref{fig:barycentric_triangles} for three representative locations: $x/H = 2$, $5$, and $7$. In both flows ($\alpha=1.0$ and 1.5, which are the predictive target of Case 1 and Case 2, respectively), the three locations correspond to the middle of the separation bubble, near the reattachment point, and after the reattachment, respectively. Overall, the Barycentric plots suggest that both random forest and neural network showed satisfactory agreement as compared to the DNS data for both Case~1 and Case~2. This is further detailed in Table~\ref{tab:errors}, which shows that the relative prediction errors of both random forest and neural network models are approximately 10\% (for $\xi$) and 5\% (for $\eta$) in the interpolation case (Case 1). These numbers are $14\%$ and $7\%$ in the extrapolation case (Case 2). Prediction percentage errors have been calculated using Frobenius norm, which for $\xi$ can be defined as: 
\begin{equation}
    \epsilon_\xi = \frac{\|\mathbf{\xi}_\text{truth} - \mathbf{\xi}_\text{predicted}\|_F}{\|\mathbf{\xi}_{truth} \|_F} 
\end{equation}
where the Frobenius norm $\|X\|_F=\sqrt{\sum_{i=1}^n\left|X_i\right|^2}$ for a vector $X \in \mathbb{R}^n$. The percentage error for $\eta$ is defined similarly. When interpreting the errors rates in Table~\ref{tab:errors}, the trends are more important than the exact numbers, because these numbers largely depends on the tuning parameters while the trends is valid in general.
The similar predictive performances of random forests and neural networks as observed above are noteworthy, because neural networks are much more complex models than random forests. Neural networks have many more tuning parameters and are much more computationally expensive to training than random forests. A benefit that comes with such a complexity is that they general perform better in extrapolations. The counter-intuitive observation here seems to suggest that the data among different training flows do not have clear intrinsic trends that can be extrapolated. In such cases, a simpler model such as random forest is the preferred choice. In general cases where neural networks are preferred, random forest can still be a valuable tool to obtain first predictions which neural networks are expected to outperform.

Another clear trend as seen from Fig.~\ref{fig:barycentric_triangles} and Table~\ref{tab:errors} is that the predictive errors in the extrapolation case (Case 2; the flow at $\alpha = 1.5$ is predicted) is much larger than in the interpolation case (Case 1, where the flow at $\alpha=1.0$ is predicted). This is expected because the case at $\alpha = 1.5$ has a much milder slope and much smaller separation bubble (see Fig.~\ref{fig:variation}b) than any of the training flows. We have performed a similar extrapolation experiment by training on $\alpha = \{0.8, 1.0, 1.2, 1.5\}$ and testing on $\alpha = 0.5$. The predictive performance is comparable to Case 1 (interpolation) above for both random forests and neural networks. This is probably because the massive separation characterizing the flow at $\alpha =0.5$ is already present in the training sets, particular in flows with $\alpha = 0.8$ and $1.0$. 
    
 \begin{table}[htbp]
\caption{Overview of the scenarios for training and testing the data-driven model. Case 1 is an interpolation in the hill slope, while Case 2 involves extrapolation in the hill slope.}
\centering
\begin{tabular}{ c c c }
\hline
 & 	Training set &	Testing set \\
\hline
Case 1 & $\alpha = \{0.5, 0.8, 1.2, 1.5\}$ & $\alpha = 1.0$  \\
Case 2 &	$\alpha = \{0.5, 0.8, 1.0, 1.2\} $ & $\alpha = 1.5$\\
\hline
\end{tabular}
\label{tab:cases}
\end{table}%

\begin{table}[htbp]
\caption{Comparison of predictive performances in error percentage of $(\xi, \eta)$ between random forests and neural networks on the interpolation and extrapolation cases.}
\centering
\begin{tabular}{ c c c }
\hline
 & 	Random Forest &	Neural Network \\
\hline
Case 1 (interpolation) & (10.4\%, 4.8\%) & (10.9\%, 4.8\%)  \\
Case 2 (extrapolation) & (13.1\%, 7.3\%)	 & (14.5\%, 6.8\%) \\
\hline
\end{tabular}
\label{tab:errors}
\end{table}%

\subsection{Demystifying machine learning based flow predictions}
\begin{figure}[htb]
    \centering
\includegraphics[width=0.45\textwidth]{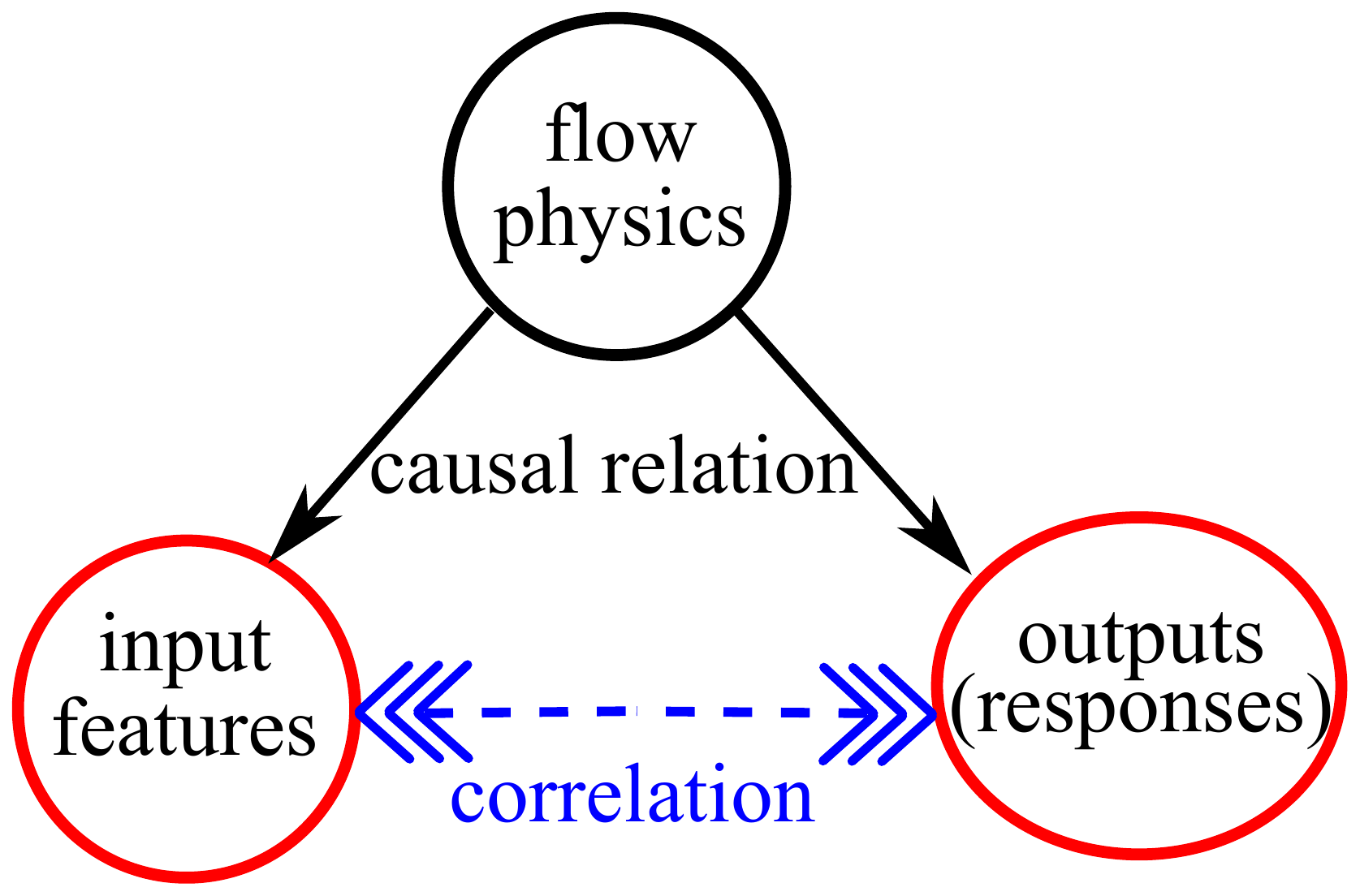}
\caption{Schematic illustration of why the machine learning based prediction works for flow problems. The input features and the output quantities stem from the same underlying physics. Machine learning algorithms discover such correlation from training data and are thus able to make physics-grounded prediction, even though the underlying physics are not explicitly identified.}
\label{fig:ml-interp}
\end{figure}

Given the good agreement between the machine learning predicted Reynolds stress anisotropy and the DNS data, it is important to examine why and how it worked. State-of-the-art machine learning methods are still based on correlations rather than causal relations. This is fundamentally different from traditional, first-principle-based modeling workflow, which typically consists of (1) identifying principles from observations, (2) formulating equations to describe such principles, and (3) solving the equations to obtain predictions. In contrast, machine learning builds upon correlation between input features and output responses. This is not to say that machine learning does not capture physics. Rather, the input and the output are manifestation of the same underlying physics and thus are inevitably correlated. Such correlations are embedded in the training data and can be discovered by machine learning algorithms. The relation between input features, output responses, and the underlying flow physics are illustrated in Fig.~\ref{fig:ml-interp}. Machine learning algorithms discover input--output correlation from training data and are thus able to make physics-grounded prediction, even though the underlying physics are not explicitly identified.

\begin{figure}[htb]
    \centering
\subfloat[Q criterion]{\includegraphics[width=0.45\textwidth]{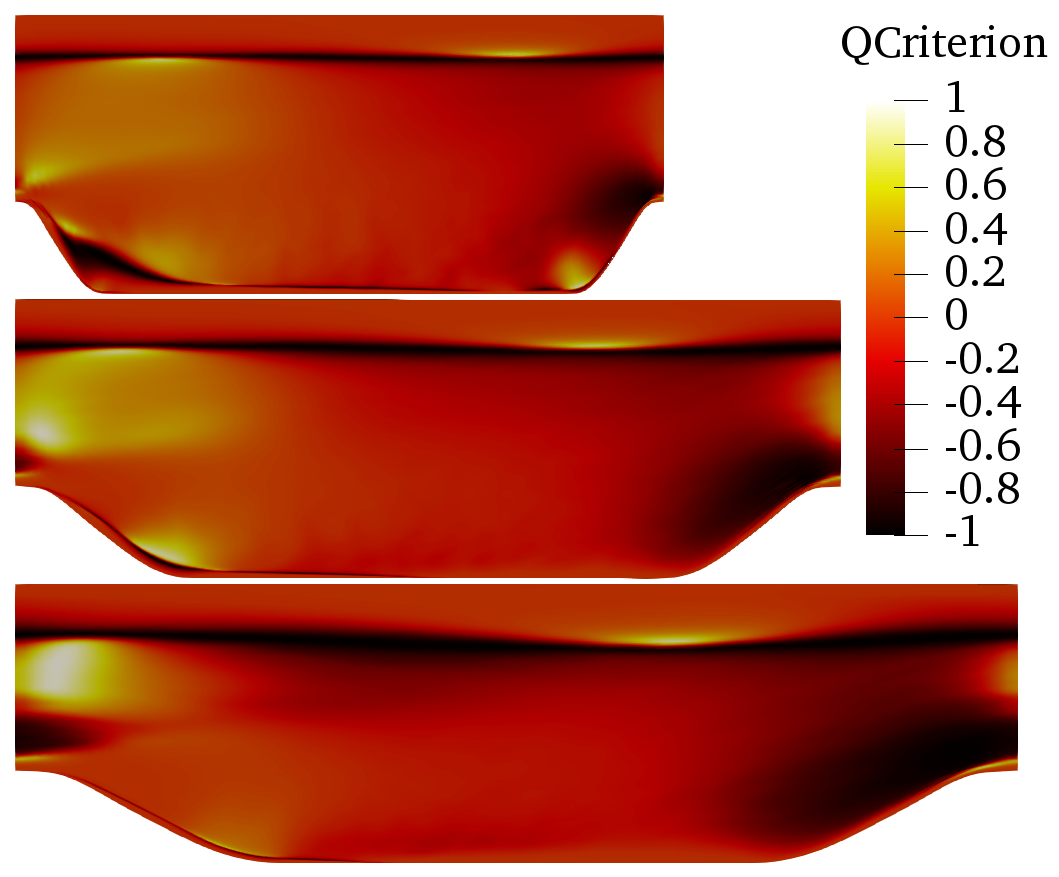}}
\subfloat[Non-orthogonality]{\includegraphics[width=0.45\textwidth]{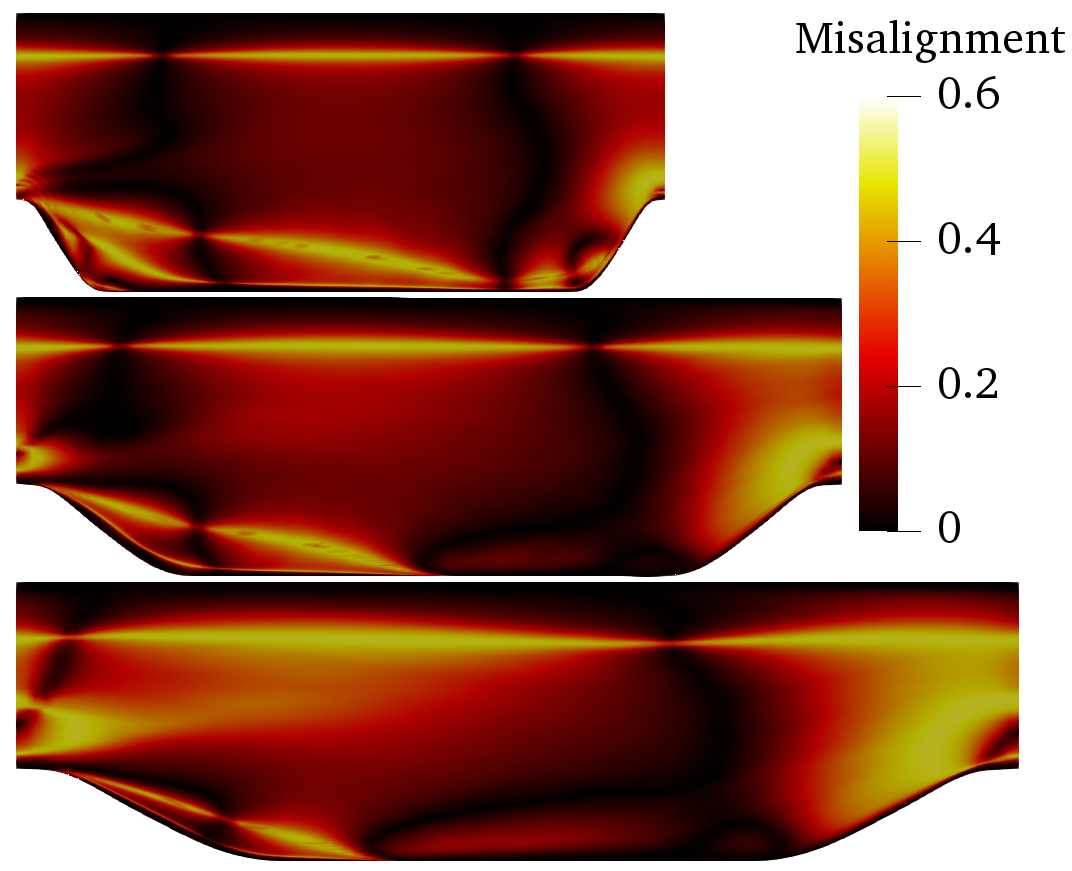}}
\caption{Color contours of selected input features, \textbf{Q criterion} ($q_1$) and \textbf{velocity/velocity-gradient non-orthogonality} ($q_7$) for three cases: $\alpha = 0.5$ (top panels), $1.0$ (middle panels), and 1.5 (bottom panels).}
\label{fig:feature-compare}
\end{figure}

The relation between input features and the flow physics are illustrated in Fig.~\ref{fig:feature-compare} by using two selected features: the Q-criterion ($q_1$) and the non-orthogonality between velocity and its gradient ($q_7$). Figure~\ref{fig:feature-compare}a shows that all three flows ($\alpha = 0.5$, $1.0$, and $1.5$) show small values (darker regions) along the upper channel and the acceleration region (near the outlet). Larger values (brighter regions) are seen near the inlet. This observation suggests that the three flows do share similar flow patterns and the Q-criterion feature reflects such similarities. Figure~\ref{fig:feature-compare}b shows that upper channels and the outer edge of the recirculation bubble are characterized by large velocity/velocity-gradient non-orthogonality, indicating their departure from parallel shear flow (channel-like flows). Again, such patterns are observed through all three flows ($\alpha = 0.5$, $1.0$, and $1.5$).

\begin{figure}[htbp]
    \centering
\subfloat[$\xi$]{\includegraphics[width=0.45\textwidth]{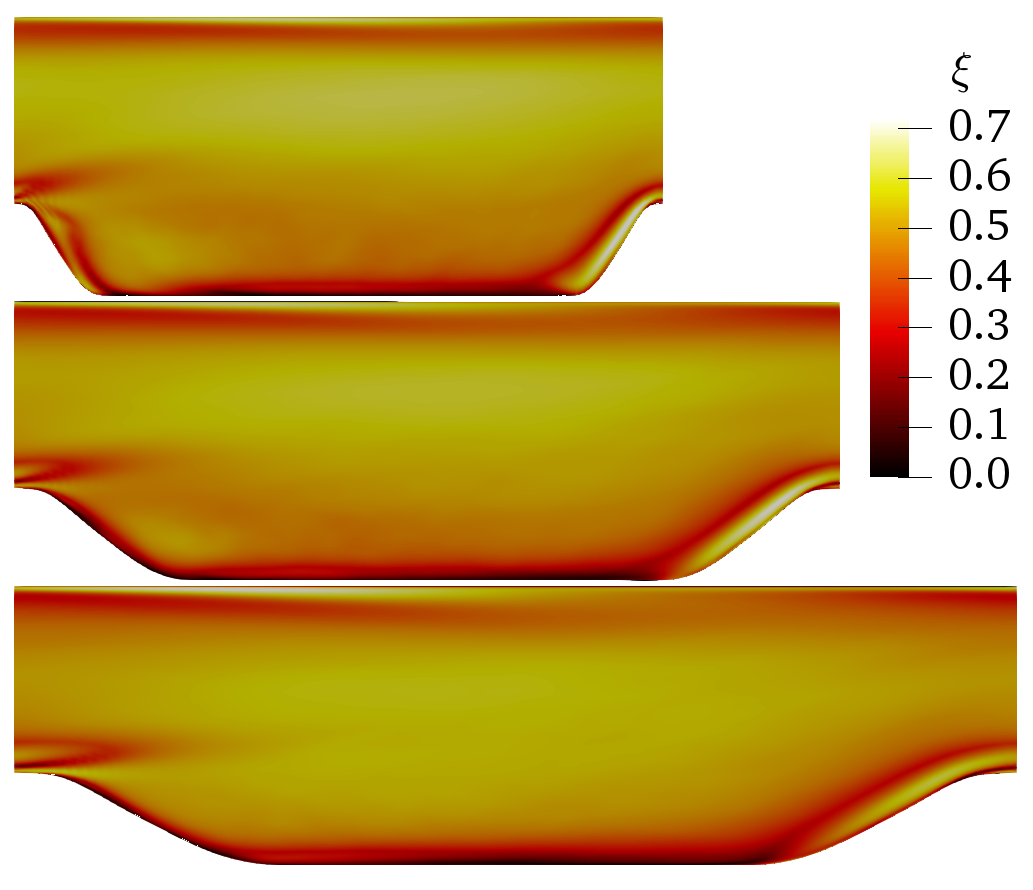}}
\subfloat[$\eta$]{\includegraphics[width=0.45\textwidth]{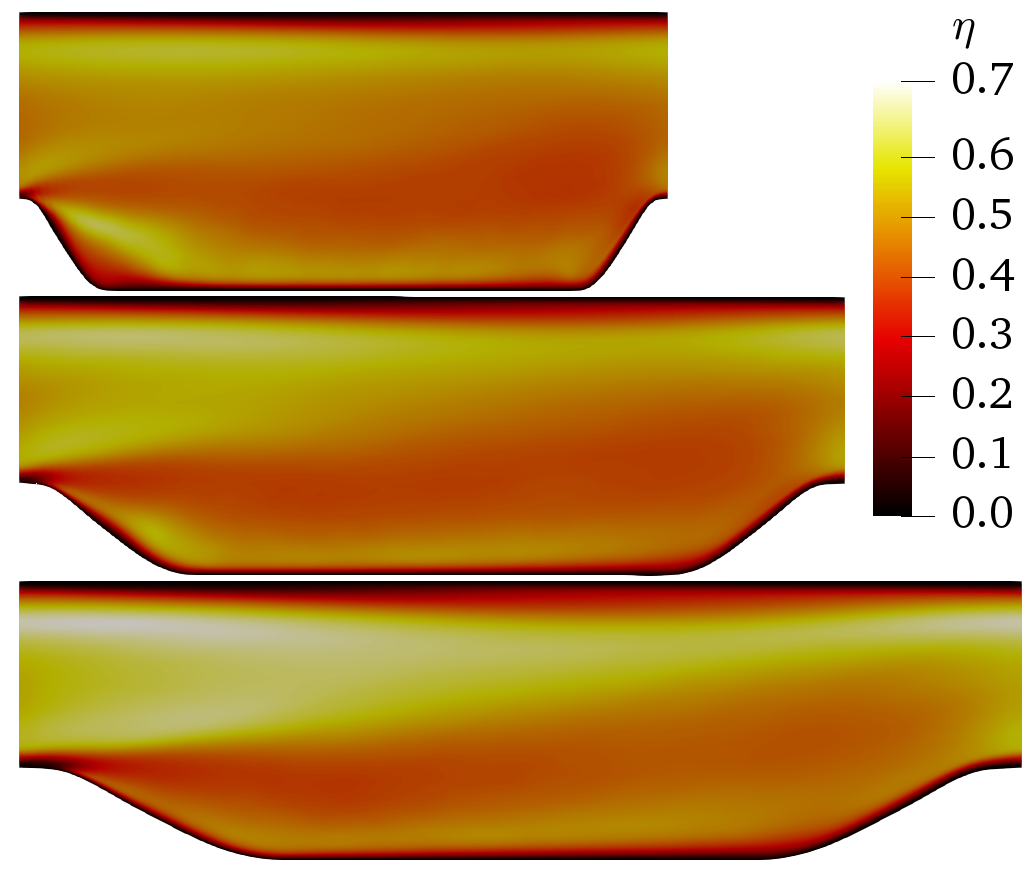}}
\caption{Plots of \textbf{anisotropy} $\xi$ and $\eta$ for three cases: $\alpha = 0.5$ (top panels), $1.0$ (middle panels), and 1.5 (bottom panels).}
\label{fig:output-compare}
\end{figure}

The output quantities also show the similar patterns across three flows as is evident from Fig.~\ref{fig:output-compare}. For example, Fig.~\ref{fig:output-compare}b shows that small values (darker regions) are found in the near-wall regions. Along with the small $ \xi$ values in the same region found from Fig.~\ref{fig:output-compare}a, this observation suggests that the turbulence states here are closer to two-component state (lower left corner of the Barycentric triangle; see Fig.~\ref{fig:bary-scheme}). In contrast, the core region downstream of the center of the inlet shows larger values of $\xi$ and $\eta$, suggesting more isotropic turbulence states.
In summary, machine learning algorithms discover the correlation between the input features and the outputs that stem from the same underlying physics to make flow predictions. The predicted trajectories in the Barycentric triangle as shown in Fig.~\ref{fig:barycentric_triangles} are in good agreement with the DNS data. This is impressive since a typical linear eddy viscosity model, which belongs to the most widely used turbulence models, are completely incapable of predicting such anisotropy correctly. Rather, it predicts an isotropic state (top vertex, 3C, in Fig.~\ref{fig:bary-scheme}) at the wall and moves towards the center of the Barycentric triangle as we trace the turbulence away from the wall.

\section{Conclusion}
\label{sec:conclude}

Data-driven turbulence modeling has emerged as a promising field in light of the long stagnation of traditional turbulence model development.  However, the development of data-driven models is hindered by the lack of datasets specifically tailored for such purposes, because existing databases are sparsely scattered in the flow parameter space and are not suitable for training and testing data-driven models. To alleviate this bottleneck, in this work we advocate the construction of benchmark datasets by systematically varying flow configurations. To this end, we present the design and generation of a dataset consisting of flow over periodic hills of various slopes. It is expected that such a dataset will be valuable for data-driven turbulence model developers to train and evaluate the predictive capabilities of their models on separated flows. We further illustrate example usage of such presented dataset in training a data-driven model that maps mean flow features to Reynolds stress anisotropy.

\appendix
\section{Description of the Hill Geometry}
\label{app:hill}
The coordinates of the first hill geometry consists of six segments of third-order polynomials described by the following equations:
\begin{align*}
\hat{y}=& \text{min}(1;1+2.42 \times 10^{-4} \hat{x}^2-7.588 \times 10^{-5} \hat{x}^3), & \hat{x} \in [0, 0.3214] \\
\hat{y}=&0.8955+3.484 \times 10^{-2} \hat{x} -3.629 \times 10^{-3} \hat{x}^2 + 6.749 \times 10^{-5} \hat{x}^3, & \hat{x} \in (0.3214, 0.5]  \\
\hat{y}=&0.9213+2.931 \times 10^{-2} \hat{x}-3.234 \times 10^{-3} \hat{x}^2+5.809 \times 10^{-5} \hat{x}^3, & \hat{x} \in (0.5, 0.7143] \\
\hat{y}=&1.445-4.927 \times 10^{-2} \hat{x} +6.95 \times 10^{-4} \hat{x}^2-7.394 \times 10^{-6} \hat{x}^3, & \hat{x} \in (0.7143, 1.071]  \\
\hat{y}=&0.6401+3.123 \times 10^{-2} \hat{x} -1.988\times 10^{-3}\hat{x}^2+2.242 \times 10^{-5} \hat{x}^3, & \hat{x} \in (1.071, 1.429] \\
\hat{y}=& \text{max}(0;2.0139-7.18 \times 10^{-2} \hat{x}+5.875 \times 10^{-4} \hat{x}^2+9.553 \times 10^{-7}\hat{x}^3), & \hat{x} \in (1.429, 1.929] 
\end{align*}
where $\hat{x} = x/H$ and $\hat{y} = y/H$ are normalized horizontal and vertical coordinates, respectively.

\newpage

\thispagestyle{empty}
\textbf{Highlights}
\begin{itemize}
    \item Identified need of data in systematically varied flows for data-driven modeling
    \item Perform DNS of flows over periodic hills with varying slopes
    \item Generated data for flows ranging from incipient to massive separations
\end{itemize}


\section*{References}

\begin{thebibliography}{10}
\expandafter\ifx\csname url\endcsname\relax
  \def\url#1{\texttt{#1}}\fi
\expandafter\ifx\csname urlprefix\endcsname\relax\def\urlprefix{URL }\fi
\expandafter\ifx\csname href\endcsname\relax
  \def\href#1#2{#2} \def\path#1{#1}\fi

\bibitem{moin97tackling}
P.~Moin, J.~Kim, Tackling turbulence with supercomputers, Scientific American
  276~(1) (1997) 46--52.

\bibitem{xiao2019quantification}
H.~Xiao, P.~Cinnella, Quantification of model uncertainty in {RANS}
  simulations: A review, Progress in Aerospace Sciences 108 (2019) 1--31.

\bibitem{launder74application}
B.~E. Launder, B.~I. Sharma, Application of the energy-dissipation model of
  turbulence to the calculation of flow near a spinning disc, Letters in Heat
  and Mass Transfer 1~(2) (1974) 131--137.

\bibitem{menter94two-equation}
F.~R. Menter, Two-equation eddy-viscosity turbulence models for engineering
  applications, AIAA Journal 32~(8) (1994) 1598--1605.

\bibitem{spalart92one}
P.~R. Spalart, S.~R. Allmaras, A one equation turbulence model for aerodynamic
  flows., AIAA Journal 94.

\bibitem{adler2012entry}
M.~Adler, M.~Wright, C.~Campbell, I.~Clark, W.~Engelund, T.~Rivellini, Entry,
  descent, and landing roadmap, NASA TA09, April.

\bibitem{national2012nasa}
{National Research Council}, {NASA} space technology roadmaps and priorities:
  Restoring {NASA}'s technological edge and paving the way for a new era in
  space, National Academies Press, 2012.

\bibitem{slotnick2014cfd}
J.~Slotnick, A.~Khodadoust, J.~Alonso, D.~Darmofal, W.~Gropp, E.~Lurie,
  D.~Mavriplis, {CFD} vision 2030 study: a path to revolutionary computational
  aerosciences, Tech. rep., National Aeronautics and Space Administration,
  Langley Research Center, Hampton, Virginia 23681-2199 (2014).

\bibitem{singh16using}
A.~P. Singh, K.~Duraisamy, Using field inversion to quantify functional errors
  in turbulence closures, Physics of Fluids 28 (2016) 045110.

\bibitem{singh16machine}
A.~P. Singh, S.~Medida, K.~Duraisamy, Machine learning-augmented predictive
  modeling of turbulent separated flows over airfoils, AIAA Journal 55~(7)
  (2017) 2215--2227.

\bibitem{wang2017physics-informed}
J.-X. Wang, J.-L. Wu, H.~Xiao, Physics-informed machine learning approach for
  reconstructing {Reynolds} stress modeling discrepancies based on {DNS} data,
  Physical Review Fluids 2~(3) (2017) 034603.

\bibitem{wu2018physics-informed}
J.-L. Wu, H.~Xiao, E.~G. Paterson, Physics-informed machine learning approach
  for augmenting turbulence models: A comprehensive framework, Physical Review
  Fluids 3 (2018) 074602.

\bibitem{ling2016reynolds}
J.~Ling, A.~Kurzawski, J.~Templeton, Reynolds averaged turbulence modelling
  using deep neural networks with embedded invariance, Journal of Fluid
  Mechanics 807 (2016) 155--166.

\bibitem{weatheritt2016novel}
J.~Weatheritt, R.~Sandberg, A novel evolutionary algorithm applied to algebraic
  modifications of the rans stress--strain relationship, Journal of
  Computational Physics 325 (2016) 22--37.

\bibitem{weatheritt2017development}
J.~Weatheritt, R.~Sandberg, The development of algebraic stress models using a
  novel evolutionary algorithm, International Journal of Heat and Fluid Flow.

\bibitem{schmelzer2019machine}
M.~Schmelzer, R.~P. Dwight, P.~Cinnella, Machine learning of algebraic stress
  models using deterministic symbolic regression, Flow, Turbulence and
  Combustion.

\bibitem{krizhevsky12imagenet}
A.~Krizhevsky, I.~Sutskever, G.~E. Hinton, Imagenet classification with deep
  convolutional neural networks, in: Advances in Neural Information Processing
  Systems, 2012, pp. 1097--1105.

\bibitem{deng09imagenet}
J.~Deng, W.~Dong, R.~Socher, L.-J. Li, K.~Li, L.~Fei-Fei, {ImageNet: A
  Large-Scale Hierarchical Image Database}, in: CVPR09, 2009.

\bibitem{kim1987turbulence}
J.~Kim, P.~Moin, R.~Moser, Turbulence statistics in fully developed channel
  flow at low reynolds number, Journal of Fluid Mechanics 177 (1987) 133--166.

\bibitem{lee15direct}
M.~Lee, R.~D. Moser, Direct numerical simulation of turbulent channel flow up
  to {Re}=5200, Journal of Fluid Mechanics 774 (2015) 395--415.

\bibitem{agard}
AGARD, A selection of test cases for the validation of {Large-Eddy Simulations}
  of turbulent flows, {AGARD} advisory report no.~345, data:
  \url{http://torroja.dmt.upm.es/turbdata/agard/}.

\bibitem{nasa-database}
C.~L. Rumsey, {NASA} {Langley} turbulence modeling portal,
  \url{https://turbmodels.larc.nasa.gov} (2018).

\bibitem{ercoftac}
ERCOFTAC, {ERCOFTAC} classic collection database,
  \url{http://cfd.mace.manchester.ac.uk/ercoftac/}.

\bibitem{jhu-database}
{Johns Hopkins University}, {Johns Hopkins} turbulence database ({JHTDB}) site,
  \url{http://turbulence.pha.jhu.edu}.

\bibitem{li08public}
Y.~Li, E.~Perlman, M.~Wan, Y.~Yang, C.~Meneveau, R.~Burns, S.~Chen, A.~Szalay,
  G.~Eyink, A public turbulence database cluster and applications to study
  {Lagrangian} evolution of velocity increments in turbulence, Journal of
  Turbulence~(9) (2008) N31.

\bibitem{pinelli10reynolds}
A.~Pinelli, M.~Uhlmann, A.~Sekimoto, G.~Kawahara, Reynolds number dependence of
  mean flow structure in square duct turbulence, Journal of fluid mechanics 644
  (2010) 107--122.

\bibitem{kti-database}
M.~Uhlmann, Fully-developed, pressure-driven flow of an incompressible,
  isothermal fluid through a straight duct with square cross section: data from
  {DNS}, \url{http://www.ifh.kit.edu/dns_data/duct/}.

\bibitem{maass96direct}
C.~Maa{\ss}, U.~Schumann, Direct numerical simulation of separated turbulent
  flow over a wavy boundary, in: Flow Simulation with High-Performance
  Computers II, Springer, 1996, pp. 227--241.

\bibitem{bentaleb2012large-eddy}
Y.~Bentaleb, S.~Lardeau, M.~A. Leschziner, Large-eddy simulation of turbulent
  boundary layer separation from a rounded step, Journal of Turbulence~(13)
  (2012) N4.

\bibitem{laval2011direct}
J.-P. Laval, M.~Marquillie, Direct numerical simulations of
  converging--diverging channel flow, in: Progress in Wall Turbulence:
  Understanding and Modeling, Springer, 2011, pp. 203--209.

\bibitem{breuer2009flow}
M.~Breuer, N.~Peller, C.~Rapp, M.~Manhart, Flow over periodic hills: Numerical
  and experimental study in a wide range of {R}eynolds numbers, Computers \&
  Fluids 38~(2) (2009) 433--457.

\bibitem{kutz17deep}
J.~N. Kutz, Deep learning in fluid dynamics, Journal of Fluid Mechanics 814
  (2017) 1--4.

\bibitem{frohlich2005highly}
J.~Fr{\"o}hlich, C.~P. Mellen, W.~Rodi, L.~Temmerman, M.~A. Leschziner, Highly
  resolved large-eddy simulation of separated flow in a channel with streamwise
  periodic constrictions, Journal of Fluid Mechanics 526 (2005) 19--66.

\bibitem{rapp2011flow}
C.~Rapp, M.~Manhart, Flow over periodic hills: an experimental study,
  Experiments in fluids 51~(1) (2011) 247--269.

\bibitem{gloerfelt2019large}
X.~Gloerfelt, P.~Cinnella, Large eddy simulation requirements for the flow over
  periodic hills, Flow, Turbulence and Combustion 103~(1) (2019) 55--91.

\bibitem{gloerfelt2019benchmark}
X.~Gloerfelt, P.~Cinnella, Benchmark database: {2D} periodic hill flow,
  \url{https://www.researchgate.net/publication/315413324_Benchmark_database_2D_periodic_hill_flow}.

\bibitem{wu2019representation}
J.-L. Wu, R.~Sun, S.~Laizet, H.~Xiao, Representation of stress tensor
  perturbations with application in machine-learning-assisted turbulence
  modeling, Computer Methods in Applied Mechanics and Engineering 346 (2019)
  707--726.

\bibitem{wu2019reynolds}
J.-L. Wu, H.~Xiao, R.~Sun, Q.~Wang, Reynolds-averaged {Navier--Stokes}
  equations with explicit data-driven {Reynolds} stress closure can be
  ill-conditioned, Journal of Fluid Mechanics 869 (2019) 553--586.

\bibitem{kravchenko&moin97}
A.~G. {\rm Kravchenko}, P.~{\rm Moin}, On the effect of numerical errors in
  large eddy simulation of turbulent flows, J. Comp. Phys. {\rm 131} (1997)
  310--322.

\bibitem{incompact3d-code}
S.~Laizet, Incompressible {Navier-Stokes} equations solver with multiple scalar
  transport equations, \url{https://github.com/xcompact3d/Incompact3d}.

\bibitem{gautier2014dns}
R.~Gautier, S.~Laizet, E.~Lamballais, A {DNS} study of jet control with
  microjets using an immersed boundary method, Int. J. of Computational Fluid
  Dynamics 28~(6-10) (2014) 393--410.

\bibitem{laizet&lamballais09}
S.~{Laizet}, E.~{Lamballais}, High-order compact schemes for incompressible
  flows: A simple and efficient method with the quasi-spectral accuracy, J.
  Comp. Phys. 228 (2009) 5989--6015.

\bibitem{laizet&li11}
S.~{Laizet}, N.~{Li}, Incompact3d: A powerfull tool to tackle turbulence
  problems with up to {O}$(10^5)$ computational cores, Int. J. Heat and Fluid
  Flow {\bf 67} (2011) 1735--1757.

\bibitem{espath2014two}
L.~Espath, L.~Pinto, S.~Laizet, J.~Silvestrini, Two-and three-dimensional
  direct numerical simulation of particle-laden gravity currents, Computers \&
  geosciences 63 (2014) 9--16.

\bibitem{schuch2018three}
F.~N. Schuch, L.~C. Pinto, J.~H. Silvestrini, S.~Laizet, Three-dimensional
  turbulence-resolving simulations of the plunge phenomenon in a tilted
  channel, Journal of Geophysical Research: Oceans 123~(7) (2018) 4820--4832.

\bibitem{lucchese2019direct}
L.~V. Lucchese, L.~R. Monteiro, E.~B.~C. Schettini, J.~H. Silvestrini, Direct
  numerical simulations of turbidity currents with evolutive deposit method,
  considering topography updates during the simulation, Computers \&
  Geosciences 133 (2019) 104306.

\bibitem{mahfoze2017skin}
O.~Mahfoze, S.~Laizet, Skin-friction drag reduction in a channel flow with
  streamwise-aligned plasma actuators, International Journal of Heat and Fluid
  Flow 66 (2017) 83--94.

\bibitem{yao2018drag}
J.~Yao, X.~Chen, F.~Hussain, Drag control in wall-bounded turbulent flows via
  spanwise opposed wall-jet forcing, Journal of Fluid Mechanics 852 (2018)
  678--709.

\bibitem{mahfoze2019reducing}
O.~Mahfoze, A.~Moody, A.~Wynn, R.~Whalley, S.~Laizet, Reducing the
  skin-friction drag of a turbulent boundary-layer flow with low-amplitude
  wall-normal blowing within a bayesian optimization framework, Physical Review
  Fluids 4~(9) (2019) 094601.

\bibitem{dairay2014turbulent}
T.~Dairay, V.~Fortun{\'e}, E.~Lamballais, L.~Brizzi, Les of a turbulent jet
  impinging on a heated wall using high-order numerical schemes, International
  Journal of Heat and Fluid Flow 50 (2014) 177--187.

\bibitem{dairay2015direct}
T.~Dairay, V.~Fortun{\'e}, E.~Lamballais, L.-E. Brizzi, Direct numerical
  simulation of a turbulent jet impinging on a heated wall, Journal of Fluid
  Mechanics 764 (2015) 362--394.

\bibitem{deskos2018development}
G.~Deskos, S.~Laizet, M.~D. Piggott, Development and validation of the
  higher-order finite-difference wind farm simulator, winc3d, in: 3rd
  international conference on renewable energies offshore (renew2018). Lisbon,
  Portugal, 2018.

\bibitem{deskos2019turbulence}
G.~Deskos, S.~Laizet, M.~D. Piggott, Turbulence-resolving simulations of wind
  turbine wakes, Renewable energy 134 (2019) 989--1002.

\bibitem{dairay2015non}
T.~Dairay, M.~Obligado, J.~C. Vassilicos, Non-equilibrium scaling laws in
  axisymmetric turbulent wakes, Journal of Fluid Mechanics 781 (2015) 166--195.

\bibitem{obligado2016nonequilibrium}
M.~Obligado, T.~Dairay, J.~C. Vassilicos, Nonequilibrium scalings of turbulent
  wakes, Physical Review Fluids 1~(4) (2016) 044409.

\bibitem{zhou2017related}
Y.~Zhou, J.~Vassilicos, Related self-similar statistics of the
  turbulent/non-turbulent interface and the turbulence dissipation, Journal of
  Fluid Mechanics 821 (2017) 440--457.

\bibitem{xiao2019data}
H.~Xiao, J.~Wu, S.~Laizet, L.~Duan, Flow over periodic hills of parameterized
  geometries: Example code and dataset for data-driven turbulence modeling, see
  also \url{https://github.com/xiaoh/para-database-for-PIML} (2019).

\bibitem{ling2015evaluation}
J.~Ling, J.~Templeton, Evaluation of machine learning algorithms for prediction
  of regions of high {Reynolds-averaged Navier--Stokes} uncertainty, Physics of
  Fluids (1994-present) 27~(8) (2015) 085103.

\bibitem{wang2019prediction}
J.-X. Wang, J.~Huang, L.~Duan, H.~Xiao, Prediction of {Reynolds} stresses in
  high-{Mach}-number turbulent boundary layers using physics-informed machine
  learning, Theoretical and Computational Fluid Dynamics 33~(1) (2019) 1--19.

\bibitem{zhu2019machine}
L.~Zhu, W.~Zhang, J.~Kou, Y.~Liu, Machine learning methods for turbulence
  modeling in subsonic flows around airfoils, Physics of Fluids 31~(1) (2019)
  015105.

\bibitem{emory2011modeling}
M.~Emory, R.~Pecnik, G.~Iaccarino, Modeling structural uncertainties in
  {Reynolds-averaged} computations of shock/boundary layer interactions, AIAA
  paper 479 (2011) 1--16.

\bibitem{banerjee2007presentation}
S.~Banerjee, R.~Krahl, F.~Durst, C.~Zenger, Presentation of anisotropy
  properties of turbulence, invariants versus eigenvalue approaches, Journal of
  Turbulence 8~(32).

\bibitem{breiman2001random}
L.~Breiman, Random forests, Machine learning 45~(1) (2001) 5--32.

\bibitem{hornik89multilayer}
K.~Hornik, M.~Stinchcombe, H.~White, Multilayer feedforward networks are
  universal approximators, Neural Networks 2~(5) (1989) 359--366.

\bibitem{friedman2001elements}
J.~Friedman, T.~Hastie, R.~Tibshirani, The elements of statistical learning,
  Springer, Berlin, 2001.

\end{thebibliography}
\end{document}